\documentstyle[aps,version2,preprint]{revtex}    
\begin{document}

\draft

\begin{title}
Ground states of integrable quantum liquids
\end{title}

\author{J.M.P. Carmelo $^{1,2}$ and N.M.R. Peres $^{2}$}
\begin{instit}
$^{1}$ Instituto de Ciencia de Materiales, C.S.I.C.,
Cantoblanco, E - 28949 Madrid, Spain
\end{instit}
\begin{instit}
$^{2}$ Department of Physics, University
of \'Evora, Apartado 94, P - 7001 \'Evora codex, Portugal
\end{instit}
\receipt{July 1994}
\begin{abstract}
Based on a recently introduced operator algebra for the
description of a class of integrable quantum liquids
we define the ground states for all canonical ensembles
of these systems. We consider the particular case of the
Hubbard chain in a magnetic field and chemical potential.
The ground states of all canonical ensembles of the model
can be generated by acting onto the electron vacuum (densities $n<1$)
or hole vacuum (densities $n>1$), suitable pseudoparticle creation
operators. We also evaluate the energy gaps of the non-lowest-weight
states (non - LWS's) and non-highest-weight states (non -
HWS's) of the eta-spin and spin algebras
relative to the corresponding ground states. For all sectors
of parameter space and symmetries the {\it exact ground state}
of the many-electron problem is in the pseudoparticle basis
the non-interacting pseudoparticle ground state. This
plays a central role in the pseudoparticle perturbation
theory.
\end{abstract}
\renewcommand{\baselinestretch}{1.656}   


\narrowtext

\section{INTRODUCTION}

In contrast to higher dimension, the many-electron
one-dimensional problem is non perturbative. However, it also
represents one of the few cases of many-body theory where
there exist a number of exact solutions available. The
most important class of such electronic systems are
the integrable models solvable by the ``Bethe ansatz'' (BA)
\cite{Bethe,Yang,Lieb,Korepinrev}. Within these integrable
``Luttinger-liquids'' \cite{Haldane,Ander} the
multicomponent models describing interacting fermions
have been extensively investigated
\cite{Yang,Lieb,Izergin,Frahm,Carmelo92,Carmelo92c,Neto,Campbell}.
In spite of the non-Fermi liquid behavior, these
systems show in the sectors of parameter space of
symmetry $U(1)$ a ``Landau-liquid'' character
\cite{Carmelo92,Carmelo92c,Neto,Campbell,Carmelo94,Carmelo95,Carmelo91}.

For some particular choices of canonical ensembles of the
Hubbard chain \cite{Lieb} and other integrable models it has been
assumed that the ground state corresponds to filling
symmetrically around the origin the BA quantum numbers
\cite{Lieb,Carmelo92c,Carmelo94,Carmelo95,Shiba,Carmelo91b,Carmelo92b}.
Following this assumption, the pseudoparticle-operator
algebra introduced in Refs. \cite{Neto,Carmelo94,Carmelo95} revealed
that, in contrast to the electronic basis, the pseudoparticle
basis is perturbative. In particular, it was found that
the natural reference state of the pseudoparticle
perturbation theory is the many-electron ground state,
which in the new basis is the non-interacting pseudoparticle
ground state. This implies that this state plays a crucial
role in the pseudoparticle perturbation theory.

In this paper we present a detailed study of the ground-state
problem which confirms that the many-electron ground
state corresponds to filling symmetrically around the
origin the BA quantum numbers and that in the
pseudoparticle basis it is the non-interacting
pseudoparticle ground state. One of our goals
is to deepen and generalize to {\it all} sectors of parameter
space of the Hubbard chain in the presence of a
magnetic field and chemical potential the study of that ground
state. Our study refers directly to the canonical ensembles
belonging to the four sectors of parameter space where the
symmetry is $U(1)\otimes U(1)$ which correspond to electronic
densities $0<n<1$ and $1<n<2$ and spin densities $-n<m<0$
and $-0<m<n$ (for $0<n<1$) and $-(2-n)<m<0$ and
$0<m<(2-n)$ (for $1<n<2$). However, it also
provides the results for the sectors of higher
symmetry and, therefore, for {\it all}
canonical ensembles of the model which correspond to
electronic densities $0\leq n\leq 2$
and spin densities $-n\leq m\leq n$ (for $0\leq n\leq 1$)
and $-(2-n)\leq m\leq (2-n)$ (for $1\leq n\leq 2$).

Previous studies of the Hubbard chain have focused mainly
on electronic densities $0\leq n\leq 1$ and spin densities
$0\leq m\leq n$. However, in spite of the fact that the
symmetries of the model provide information for the
values of the physical quantities at densities $1\leq
n\leq 2$ and spin densities $-(2-n)\leq m\leq (2-n)$
and densities $0\leq n\leq 1$ and spin densities
$-n\leq m\leq 0$ from the corresponding
values at densities $0\leq n\leq 1$ and spin
densities $0\leq m\leq n$, we show in this paper that
from the point of view of the pseudoparticle operator algebra it is
useful to consider explicitly all parameter-space sectors.

In the case of integrable models of simple Abelian $U(1)$
symmetry the elementary excitations are generated by a
single type of pseudoparticles \cite{Anto}. In
this paper we find that the description of the
low-energy excitations of non-Abelian integrable
Hamiltonians involves a larger set of pseudoparticles.
We show that the description of all gapless excitations
branches of the four $U(1)\otimes U(1)$ sectors of
parameter space of the Hubbard chain in a
magnetic field and chemical potential involves
eight types of pseudoparticles. The corresponding
pseudoparticle algebra generates the low-energy Hamiltonian
eigenstates from the electron vacuum
(densities $0<n<1$) and hole vacuum (densities $1<n<2$),
respectively. The related study of the low-energy physics
of the sectors of parameter space of higher symmetry
in terms of the above pseudoparticles will
be presented elsewhere \cite{Nuno95}. This study
leads to the symmetry transformations of the eight types
of pseudoparticles and reveals that the holons, antiholons,
and two types of spinons of Ref. \cite{Essler} are
nothing but limiting cases of our general pseudoparticles.

In the pseudoparticle basis the ground state of the
many-body problem is, for {\it all}
canonical ensembles and corresponding symmetries, a
``non-interacting'' state, {\it i.e.}, a simple Slater determinant
of filled pseudoparticle levels. In addition, in the above
sectors of symmetry $U(1)\otimes U(1)$
the elementary excitations simply correspond to
pseudoparticle-pseudohole processes around
this ground state.

The present study of the low-energy physics of the
four sectors of symmetry $U(1)\otimes U(1)$ of
the Hubbard chain in a magnetic field and
chemical potential introduces the eight types
of pseudoparticles needed for a deeper
understanding and description of the gapless
excitations in the sectors of higher symmetry
$SO(4)$, $SU(2)\otimes U(1)$, and $U(1)\otimes SU(2)$.
Therefore, the present study is a necessary step
for the full characterization and understanding
of the low-energy physics of the integrable non-Abelian
many-electron quantum problems.

In Section II we generalize the perturbation theory
introduced in Ref. \cite{Neto,Carmelo94} to all the
$U(1)\otimes U(1)$ sectors of parameter space of
the model. This requires the introduction of eight
pseudoparticle branches.

In Sec. III we confirm that in the case of canonical
ensembles belonging the sectors of parameter space of
lowest symmetry $U(1)\otimes U(1)$ the ground state corresponds
to filling symmetrically around the origin the BA quantum numbers.

In Section IV we show that in all canonical ensembles of
the four $U(1)\otimes U(1)$ sectors of parameter space the
non-LWS's and non-HWS's of the eta-spin and spin algebras have
an energy gap relative to the corresponding ground state.
In addition, we evaluate the smallest of these gaps.

Finally, in Sec. V we present the discussion and concluding
remarks.

\section{PSEUDOPARTICLE BRANCHES OF THE FOUR $U(1)\otimes U(1)$ SECTORS}

Consider the Hamiltonian describing the Hubbard chain
in a magnetic field $H$ and chemical potential
$\mu$ \cite{Frahm,Carmelo92c,Carmelo91b,Carmelo92b}:

\begin{equation}
\hat{H} = \hat{H}_{SO(4)} + 2\mu\hat{\eta }_z +
2\mu_0 H\hat{S}_z \, ,
\label{Hamilt}
\end{equation}
where

\begin{equation}
\hat{H}_{SO(4)} = -t\sum_{j,\sigma}
\left[c_{j\sigma}^{\dag }c_{j+1\sigma} +
c_{j+1\sigma}^{\dag }c_{j\sigma}\right] +
U\sum_{j} [c_{j\uparrow}^{\dag }c_{j\uparrow} - 1/2]
[c_{j\downarrow}^{\dag }c_{j\downarrow} - 1/2] \, .
\label{Hamilt}
\end{equation}
Here the operator $c_{j\sigma}^{\dagger}$ ($c_{j\sigma}$)
creates (annihilates) one spin $\sigma$ electron at the
site $j$ and $t$, $U$, $\mu$, $H$, and $\mu _0$ are the transfer
integral, the onsite Coulomb interaction, the chemical
potential, the magnetic field, and the Bohr magneton,
respectively. The operators,

\begin{equation}
\hat{\eta}_z = -{1\over 2}[N_a - \sum_{\sigma}\hat{N}_{\sigma }] \, ,
\hspace{1cm}
\hat{S}_z = -{1\over 2}\sum_{\sigma}\sigma\hat{N}_{\sigma } \, ,
\end{equation}
are the diagonal generators of the $SU(2)$ eta-spin
and spin algebras, respectively \cite{Yang89,Korepin}.
$\sigma$ refers to the spin projections $\sigma =\uparrow\,
,\downarrow$ when used as an operator or function index
and is given by $\sigma =\pm 1$ otherwise. In Eq. $(3)$
$\hat{N}_{\sigma }=\sum_{j} c_{j\sigma }^{\dagger }c_{j\sigma }$
is the number operator for $\sigma$ spin-projection
electrons.

The model $(1)-(2)$ describes an interacting quantum system
of $N_{\uparrow}$ up-spin electrons and $N_{\downarrow}$
down-spin electrons on a chain of $N_a$ sites with lattice
constant $a$. We use periodic boundary conditions and
consider $N_a$ to be even and, when $N=N_a$ (half filling),
both $N_{\uparrow}$ and $N_{\downarrow}$ to be odd.
Henceforth we employ units such that $a=t=\mu_0=\hbar =1$.
The Fermi momenta are given by $k_{F\sigma}=\pi n_{\sigma}$
and $k_F=[k_{F\uparrow}+k_{F\downarrow}]/2=\pi n/2$, where
$n_{\sigma}=N_{\sigma}/N_a$ and $n=N/N_a$. The
dimensionless onsite interaction is $u=U/4t$.

In the absence of the chemical-potential and magnetic-field
terms the Hamiltonian $(1)$ reduces to $(2)$ and
has $SO(4) = SU(2) \otimes SU(2)/Z_2$ symmetry
\cite{Yang89,Korepin,Heilmann,Lieb89,Zhang,Ostlund}.
Since $N_a$ is even, the operator
${\hat{\eta}}_z+{\hat{S}}_z$ (see Eq. $(3)$) has only integer
eigenvalues and all half-odd integer representations of
$SU(2) \otimes SU(2)$ are projected out
\cite{Essler,Korepin}. The two $SU(2)$ algebras -- eta spin
and spin  -- have diagonal generators given by Eq. (3) and
off-diagonal generators \cite{Essler,Korepin}

\begin{equation}
\hat{\eta} = \sum_{j} (-1)^j c_{j\uparrow}c_{j\downarrow}
\, , \hspace{1cm}
\hat{\eta}^{\dagger} = \sum_{j} (-1)^j c^{\dagger }_{j\downarrow}
c^{\dagger }_{j\uparrow} \, ,
\end{equation}
and

\begin{equation}
\hat{S} = \sum_{j} c^{\dagger }_{j\uparrow}c_{j\downarrow}
\, , \hspace{1cm}
\hat{S}^{\dagger} = \sum_{j} c^{\dagger }_{j\downarrow}
c_{j\uparrow} \, ,
\end{equation}
respectively.

In the presence of both the magnetic field and chemical
potential terms, the symmetry is reduced to $U(1)\otimes U(1)$,
with $\hat{\eta}_z$ and $\hat{S}_z$ commuting with $\hat{H}$.
The eigenvalues ${\eta}_z$ and $S_z$ are determined by the
values of the conserved numbers, as shown by Eq. $(3)$.
According to these eigenvalues, the system has different
symmetries as follows \cite{Carmelo92c,Carmelo94}: when
$\eta_z\neq 0$ and $S_z\neq 0$ the symmetry is
$U(1)\otimes U(1)$, for $\eta_z= 0$ and $S_z\neq 0$ (and
$\mu =0$) it is $SU(2)\otimes U(1)$, when $\eta_z\neq 0$
and $S_z= 0$ it is $U(1)\otimes SU(2)$, and at $\eta_z= 0$
and $S_z= 0$ (and $\mu =0$) the symmetry is $SO(4)$. The
$U(1)\otimes U(1)$ symmetry sectors {\it always} correspond
to two non-zero eigenvalues of the diagonal generators,
whereas in sectors of higher symmetry, one or both of these
eigenvalues vanish. There are four $U(1)\otimes U(1)$ sectors
of parameter space which correspond to $\eta_z< 0$ and $S_z< 0$;
$\eta_z< 0$ and $S_z> 0$; $\eta_z> 0$ and $S_z< 0$; and
$\eta_z> 0$ and $S_z> 0$. There are two $SU(2)\otimes U(1)$
[and $U(1)\otimes SU(2)$] sectors of parameter
space which correspond to $S_z< 0$ and $S_z> 0$
(and to $\eta_z< 0$ and $\eta_z> 0$). There is one $SO(4)$
sector of parameter space [which is the above
$\eta_z= 0$ (and $\mu =0$) and $S_z= 0$ ``point''].

We restrict our analysis to the four sectors
of lowest symmetry $U(1)\otimes U(1)$. In this case
we can define the parameters $l=sgn (\eta_z)1$ and $l'=sgn
(S_z)1$ which classify these four sectors: we denote
each of them by $(l,l')$ sector. The
$(-1,-1)$; $(-1,1)$; $(1,-1)$; and $(1,1)$ sectors
refer to electronic densities and spin densities $0<n<1$ and
$0<m<n$; $0<n<1$ and $-n<m<0$; $1<n<2$ and $0<m<(2-n)$;
and $1<n<2$ and $-(2-n)<m<0$, respectively.

In the $(-1,-1)$ sector the BA solution
refers only to the LWS's of both the eta-spin
and spin algebras. Note that the Hamiltonian
eigenstates of the remaining three $(l,l')$
sectors can be generated by multiple application
of the operators $\hat{\eta }^{\dag }$ $(4)$ and
$\hat{S}^{\dag }$ $(5)$ to the LWS's of the $(-1,-1)$
sector \cite{Yang,Korepin}. We find below
that in the sectors where $\eta_z >0$
and (or ) $S_z>0$ the ground states and low-energy
Hamiltonian eigenstates are HWS's of the eta-spin
and (or ) spin algebras. However, the generation
of these HWS's from the corresponding LWS's
with the same $\eta $ and $S$ values [and $\eta_z =-\eta_z$
and (or ) $S_z =-S_z$] represents a too complicated
and indirect description of the low-energy physics
of the $\eta_z >0$ and (or ) $S_z>0$ sectors
of parameter space. Instead, in this paper we describe
the low-energy physics of these sectors directly from
the corresponding BA solutions. We emphasize that the
set of LWS's and (or ) HWS's of each of the four
$(l,l')$ sectors of symmetry $U(1)\otimes U(1)$
represent four alternative choices for starting
states to construct a complete set of Hamiltonian
eigenstates [see Ref. \cite{Korepin} where the
LWS's of the $(-1,-1)$ sector are used as starting
states].

Since in the $(1,-1)$ [and $(-1,1)$] sector the
low-energy physics is determined by HWS's [and
LWS's] of the eta-spin algebra and LWS's [and
HWS's] of the spin algebra, henceforth we refer the
LWS's and (or ) HWS's of the four sectors as [LWS,LWS]'s [sector
$(-1,-1)$], [LWS,HWS]'s  [sector $(-1,1)$], [HWS,LWS]'s  [sector
$(1,-1)$], and [HWS,HWS]'s  [sector $(1,1)$].
(This notation refers to the order [eta-spin,spin].)

Let $|\eta_z,S_z\rangle$ be a regular BA Hamiltonian
eigenstate corresponding to the canonical ensemble of
eigenvalues $\eta_z$ and $S_z$. By regular BA Hamiltonian
eigenstates we mean here these states to which the
BA solution of any of the four $U(1)\otimes U(1)$
sectors directly refers. Such states can
be [LWS,LWS]'s, [LWS,HWS]'s, [HWS,LWS]'s, or [HWS,HWS]'s
according to the sector of parameter space.
The BA solution for the sectors $(-1,1)$, $(1,-1)$,
and $(1,1)$ is obtained as for the usual sector
$(-1,-1)$, except that instead of the equations

\begin{equation}
\hat{\eta}|\eta_z,S_z\rangle =
\hat{S}|\eta_z,S_z\rangle = 0 \, ,
\hspace{0.5cm} for \hspace{0.5cm} \eta_z<0\, , S_z<0 \, ,
\end{equation}
which refer to [LWS,LWS]'s, the following equations have
to be fulfilled

\begin{equation}
\hat{\eta}|\eta_z,S_z\rangle =
\hat{S}^{\dag}|\eta_z,S_z\rangle = 0 \, ,
\hspace{0.5cm} for \hspace{0.5cm}  \eta_z<0\, , S_z>0 \, ,
\end{equation}
for the $(-1,1)$ sector where the regular BA
Hamiltonian eigenstates are [LWS,HWS]'s;

\begin{equation}
\hat{\eta}^{\dag}|\eta_z,S_z\rangle =
\hat{S}|\eta_z,S_z\rangle = 0 \, ,
\hspace{0.5cm} for \hspace{0.5cm}  \eta_z>0\, , S_z<0 \, ,
\end{equation}
for the $(1,-1)$ sector where the regular BA
Hamiltonian eigenstates are [HWS,LWS]'s; and

\begin{equation}
\hat{\eta}^{\dag}|\eta_z,S_z\rangle =
\hat{S}^{\dag}|\eta_z,S_z\rangle = 0 \, ,
\hspace{0.5cm} for \hspace{0.5cm} \eta_z>0\, , S_z>0 \, ,
\end{equation}
for the $(1,1)$ sector where the regular BA Hamiltonian
eigenstates are [HWS,HWS]'s.

In References \cite{Neto,Carmelo94,Carmelo95} the LWS's of the usual
$(-1,-1)$ sector were devided into two types: the LWS's I,
which refer to real BA rapidities;
and the LWS's II. Some or all the rapidities which
describe the latter states are complex and
non real \cite{Carmelo94}. In the canonical
ensembles of that $U(1)\otimes U(1)$ sector both the
non-LWS's and the LWS's II have finite energy gaps relative
to the ground state \cite{Carmelo94}.

As in the case of the [LWS,LWS]'s of the $(-1,-1)$ sector,
the [LWS,HWS]'s, [HWS,LWS]'s, and [HWS,HWS]'s
can be devided into [LWS,HWS]'s I, [HWS,LWS]'s I, and
[HWS,HWS]'s I, described by real rapidities,
and [LWS,HWS]'s II, [HWS,LWS]'s II, and [HWS,HWS]'s II.
Some or all the rapidites which describe the latter
states are complex and non real. In the four $(l,l')$ sectors
the non-LWS's and non-HWS's and the states II have an
energy gap relative to the ground state, which is always a
[LWS,LWS] I, [LWS,HWS] I, [HWS,LWS] I, or [HWS,HWS] I. In
Sec. IV we will show that for the non-LWS's and non-HWS's,
whereas the gaps of the states II will be evaluated in a
following paper \cite{Nuno94}. At energy ranges smaller than those gaps
the  Hilbert subspace spanned by the regular BA states I
with common $\eta_z$ and $S_z$ eigenvalues
{\it coincides} with the full Hilbert space of the
quantum problem (spanned by states with these eigenvalues).
The perturbation theory introduced
in Refs. \cite{Neto,Carmelo94} for the $(-1,-1)$
sector is generalized here for all $(l,l')$ sectors and
refers to that Hilbert space. It corresponds to energy
scales {\it smaller} than the above gaps.

Henceforth, we will call state I
any [LWS,LWS] I, [LWS,HWS] I, [HWS,LWS] I, or [HWS,HWS] I, and
state II  any [LWS,LWS] II, [LWS,HWS] II, [HWS,LWS]
II, or [HWS,HWS] II. We note that a state I or a state
II is always a regular BA Hamiltonian eigenstate
and, therefore, is not, by definition, a non-LWS and (or )
non-HWS.

Let us generalize to the states I of all the
$U(1)\otimes U(1)$ sectors the pseudoparticle-operator
algebra of the $(-1,-1)$ sector. In each of the $(l,l')$
sectors we have two types of pseudoparticles.
Combining the four sectors we have eight types of operators
$b^{\dag }_{q\alpha(l,l')}$ and $b_{q\alpha(l,l')}$ which
obey the usual fermionic algebra

\begin{equation}
\{b^{\dag }_{q\alpha(l,l')},b_{q'\alpha'(l'',l''')}\}
= \delta_{q,q'}\delta_{\alpha ,\alpha'}\delta_{l,l''}
\delta_{l',l'''} \, ,
\end{equation}
and

\begin{equation}
\{b^{\dag }_{q\alpha(l,l')},b^{\dag }_{q'\alpha'(l'',l''')}\}
= 0 \, ,\hspace{2cm}
\{b_{q\alpha(l,l')},b_{q'\alpha'(l'',l''')}\}
= 0 \, .
\end{equation}
Therefore, there are eight quantum numbers or
``colors'' $\alpha(l,l')$, which are $c(-1,-1)$, $c(1,-1)$,
$c(-1,1)$, $c(1,1)$, $s(-1,-1)$, $s(1,-1)$, $s(-1,1)$, and
$s(1,1)$. These correspond to eight different branches
of $\alpha (l,l')$ pseudoparticles.

In each $(l,l')$ $U(1)\otimes U(1)$ sector only the
corresponding $\alpha (l,l')$ pseudoparticles participate in
the construction of the states I, as we discuss below.
Therefore, the algebra of Eqs. $(10)$ and
$(11)$ may be replaced by $\{b^{\dag }_{q\alpha(l,l')},b_{q'\alpha'(l,l')}\}
= \delta_{q,q'}\delta_{\alpha ,\alpha'}$ and
$\{b^{\dag }_{q\alpha(l,l')},b^{\dag }_{q'\alpha'(l,l')}\}
= 0$, respectively. The corresponding discrete pseudomomentum values
are

\begin{equation}
q_j={2\pi\over
{N_a}}I_j^{\alpha(l,l')} \, ,
\end{equation}
where $I_j^{\alpha(l,l')}$ are {\it consecutive} integers
or half integers. The representation $(12)$ was used by Yang
and Yang in the case of the one-dimensional boson gas with
repulsive delta-function interaction \cite{Yang69}.

There are $N_{\alpha (l,l')}^*$ possible
$I_j^{\alpha (l,l')}$ values. A state I is
specified by the distribution of $N_{\alpha(l,l')}$
occupied values, which we call $\alpha(l,l')$
pseudoparticles, over the $N_{\alpha(l,l')}^*$
available values.

Since only single and zero occupancy of
the values $I_j^{\alpha (l,l')}$ are allowed,
only pseudoparticles of the color $\alpha (l,l')$ can occupy
the states labeled by the numbers $I_j^{\alpha (l,l')}$. Therefore,
the pseudoparticles have a fermionic character, as assured by
the anticommuting algebra $(10)$ and $(11)$. (The BA wave function vanishes
for configurations showing other than single and zero occupancy
of the values $I_j^{\alpha (l,l')}$ \cite{Korepinrev,Carmelo94}.)
For the Hubbard model the BA spatial wave function
for the states I depends on the quantum numbers
$I_j^{\alpha (l,l')}$ through two sets of real numbers,
which many authors call rapidities.

The BA solution for the model $(1)$ can be obtained
directly for all sectors of parameter space.
In Appendix A we provide a short presentation of that solution
in the case of the present four $(l,l')$ sectors of symmetry
$U(1)\otimes U(1)$. In that Appendix we introduce
the rapidity functions, two-pseudoparticle phase
shifts, and some other quantities needed in this paper.
Equations (A1) and (A2) define the above rapidites for a given
choice of the distribution occupancy of the numbers
$I_j^{\alpha (l,l')}$, i.e. for a given state I. (For the
states II some or all rapidities are complex, non-real,
numbers \cite{Nuno94}.)

There are $N_{\alpha(l,l')}^*-N_{\alpha(l,l')}$ empty values, which
we call $\alpha(l,l')$ pseudoholes.
Carrying the BA solution in the same manner as for the
usual $(-1,-1)$ sector we find the numbers $N_{\alpha(l,l')}^*$
and $N_{\alpha(l,l')}$ for the remaining three sectors.
In Table 1 we give these numbers for the four sectors.
These are conserving numbers involving the numbers of
lattice sites $N_a$ and of $\sigma$ electrons $N_{\sigma}$.

The eta spin $\eta$ and spin $S$ values [and the
corresponding eigenvalues of the diagonal generators $(3)$,
$\eta_z=l \eta$ and $S_z=l' S$] can be expressed in terms
of the numbers of pseudoholes as follows

\begin{equation}
\eta = {1\over 2}[N_{c(l,l')}^*-N_{c(l,l')}]
\, ,
\end{equation}
and

\begin{equation}
S = {1\over 2}[N_{s(l,l')}^*-N_{s(l,l')}]
= {1\over 2}[N_{c(l,l')}-2N_{s(l,l')}] \, ,
\end{equation}
respectively. The numbers $I_j^{c(l,l')}$ are integers
(or half integers) for $N_{s(l,l')}$ even (or odd), and
$I_j^{s(l,l')}$ are integers (or half integers) for $N_{s(l,l')}^*$
odd (or even).

Let $|V;-1,\mp1\rangle$ and $|V;1,\mp1\rangle$ be the
vacuum states which correspond to taking the limit
of the electronic density $n\rightarrow 0$ and
$n\rightarrow 2$, respectively, keeping $\pm
N_{\uparrow}>\pm N_{\downarrow}$. In $|V;-1,\mp1\rangle$
and $|V;1,\mp1\rangle$ there are
only holes and electrons, respectively.

Since the colors $\alpha (l,l')$ and the pseudomomentum $q$ are
the only quantum numbers involved in the description of the
pseudoparticles whose occupancy configurations define the
states I, all the corresponding allowed configurations can be
generated by applying to the corresponding vacuum $|V;l,l'
\rangle $ the $\sum_{\alpha}N_{\alpha (l,l')}$ creation
operators $b^{\dag }_{q\alpha(l,l')}$, i.e.

\begin{equation}
|\eta_z,S_z\rangle = \prod_{\alpha=c,s}
[\prod_{q_j=q_1 }^{q_{N_{\alpha (l,l')}}}
b^{\dag }_{q\alpha (l,l')}]
|V;l,l'\rangle \, ,
\end{equation}
[here $l=sgn (\eta_z)$, $l'=sgn (S_z)$] where
the set of $q_j$ values, $q_1, q_2,...,
q_{N_{\alpha (l,l')}}$, refer to the $N_{\alpha (l,l')}$
occupied pseudomomenta out of the $N_{\alpha (l,l')}^*$ available
pseudomomentum values.

In each canonical ensemble belonging the $(l,l')$ sector,
out of the total of regular BA Hamiltonian eigenstates
I and II,

\begin{equation}
\left[
\begin{array}{c}
N_a\\
N_{s(l,l')}^*
\end{array}
\right]\left[
\left[
\begin{array}{c}
N_a\\
N_{s(l,l')}
\end{array}
\right]+
\left[
\begin{array}{c}
N_a\\
N_{s(l,l')} - 2
\end{array}
\right]\right] - \left[
\left[
\begin{array}{c}
N_a\\
N_{s(l,l')}^* + 1
\end{array}
\right] +
\left[
\begin{array}{c}
N_a\\
N_{s(l,l')}^* - 1
\end{array}
\right]\right]
\left[
\begin{array}{c}
N_a\\
N_{s(l,l')} - 1
\end{array}
\right]
\end{equation}
(this formula is a generalization of the $(-1,-1)$ result
of Ref. \cite{Korepin}), there are

\begin{equation}
\left[
\begin{array}{c}
N_a\\
N_{s(l,l')} + N_{s(l,l')}^*
\end{array}
\right]
\left[
\begin{array}{c}
N_{s(l,l')}^*\\
N_{s(l,l')}
\end{array}
\right]
\end{equation}
Hamiltonian eigenstates I. (The square brackets in the above
equation refer to the usual combinatoric coefficents.)

We emphasize that in the suitable limits formula
$(17)$ also applies to canonical ensembles of symmetry
$SO(4)$, $SU(2)\otimes U(1)$, and $U(1)\otimes SU(2)$.
For instance, in the $n=1$ and $m=0$ $SO(4)$ canonical ensemble
for any of the four choices of $(l,l')$ numbers
Eq. $(17)$ gives one. In this case these four choices
are alternative representations of the same canonical
ensemble, i.e. when $\eta_z =0$ (and $\mu =0$) and
$S_z  =0$ there is only one state
I with $\eta =S=0$. This is both a LWS and a HWS of
the eta-spin and spin algebras and is the $SO(4)$ ground
state \cite{Nuno95}. In this canonical ensemble
there are neither LWS's I nor HWS's I excited
singlet states of the eta-spin and (or ) spin algebras.

In each canonical ensemble of eigenvalues $\eta_z$ and
$S_z$ the states I of the form $(15)$ and of total number given
by Eq. $(17)$ constitute a complete orthonormal basis which
spans an important Hilbert subspace, which we call
$\cal {H}_I$. At energy scales smaller than the gaps for
non-LWS's, non-HWS's,  and states II, ${\cal H}_I$ represents
the full accessible Hilbert space. We emphasize that, in general,
many states I in ${\cal H}_I$ have energies larger than these
gaps. The low-energy physics is determined only
by the states I whose energies are smaller than
such gaps.

In Ref. \cite{Carmelo94}, and following Refs.
\cite{Lieb,Carmelo92c,Neto,Carmelo91b,Carmelo92b}, it was
assumed for the case of the sector $(-1,-1)$ that out of all
corresponding states I of form $(15)$,
the ground state of eigenvalues $\eta_z$ and $S_z$
corresponds to filling symmetrically around the
origin $N_{\alpha (-1,-1)}$ consecutive
$I_j^{\alpha (-1,-1)}$ values of all colors
$\alpha (-1,-1)$. In the present general case
this leads to

\begin{equation}
|0;\eta_z,S_z\rangle = \prod_{\alpha=c,s}
[\prod_{q=q_{F\alpha (l,l')}^{(-)}}^{q_{F\alpha
(l,l')}^{(+)}} b^{\dag }_{q\alpha(l,l')}]
|V;l,l'\rangle \, ,
\end{equation}
where when $N_{\alpha(l,l')}$ is odd (or even) and
$I_j^{\alpha(l,l')}$ are integers (or half
integers) the pseudo-Fermi points are symmetric
and read

\begin{equation}
q_{F\alpha (l,l')}^{(+)}=-q_{F\alpha (l,l')}^{(-)}
={\pi\over {N_a}}[N_{\alpha(l,l')}-1] \, .
\end{equation}

If all pseudo-Fermi points are symmetric the
state $(18)$ has zero momentum and is
nondegenerate. On the other hand, when at least one of the
pseudoparticle or antipseudoparticle numbers $N_{\alpha(l,l')}$
is odd (or even) and $I_j^{\alpha(l,l')}$ are half integers (or
integers) the corresponding pseudo-Fermi points are nonsymmetric
and read either

\begin{equation}
q_{F\alpha (l,l')}^{(+)} =
{\pi\over {N_a}}N_{\alpha(l,l')} \, , \hspace{2cm}
-q_{F\alpha (l,l')}^{(-)}={\pi\over
{N_a}}[N_{\alpha(l,l')}-2] \, ,
\end{equation}
or

\begin{equation}
q_{F\alpha (l,l')}^{(+)}={\pi\over
{N_a}}[N_{\alpha(l,l')}-2] \, , \hspace{2cm}
-q_{F\alpha (l,l')}^{(-)} =
{\pi\over {N_a}}N_{\alpha(l,l')} \, .
\end{equation}
In this case the state $(18)$ has finite momentum
and is degenerate. Equivalent expressions can be obtained
for the limits of the pseudo-Brioullin zones,
$q_{\alpha (l,l')}^{(\pm)}$,
if we replace $N_{\alpha(l,l')}$ by $N_{\alpha(l,l')}^*$
in Eqs. $(19)-(21)$. Except for terms of order $1/N_a$, we
have that $q_{F\alpha (l,l')}^{(+)}=-q_{F\alpha (l,l')}^{(-)}
=q_{F\alpha (l,l')}$ and $q_{\alpha (l,l')}^{(+)}=
-q_{\alpha (l,l')}^{(-)}=q_{\alpha (l,l')}$, where

\begin{equation}
q_{F\alpha (l,l')} = {\pi N_{\alpha(l,l')}\over {N_a}}
\, , \hspace{1cm} q_{\alpha (l,l')} =
{\pi N_{\alpha(l,l')}^*\over {N_a}} \, .
\end{equation}

In some studies the full expressions $(19)-(21)$
have to be used because the terms of order
$1/N_a$ play an important role. On the
other hand, many quantities are in the thermodynamic
limit insensitive to these $1/N_a$ corrections
and we can replace $q_{F\alpha (l,l')}^{(\pm)}$
in many expressions by the pseudomomenta $\pm q_{F\alpha
(l,l')}$ $(22)$. In Table 1 we present the values of the
pseudo-Fermi points and limits of the pseudo-Brillouin
zones $(22)$ for the four $(l,l')$ sectors.

In Sec. III we will confirm that the ground state of
canonical ensembles of sectors of parameter space
of symmetry $U(1)\otimes U(1)$ has the form $(18)$.

The Hamiltonian eigenstates $(15)$, in number of $(17)$ and of
which the state $(18)$ represents a particular choice,
can be rewritten relatively to the latter state.
In this case they correspond to
pseudoparticle-pseudohole processes around the reference
configuration $(18)$ and read

\begin{equation}
|\eta_z,S_z\rangle = \prod_{\alpha=c,s}
[\prod_{i,j=1}^{N_{ph}^{\alpha(l,l')}}
b^{\dag }_{q_j\alpha(l,l')}b_{q_i\alpha(l,l')}]
|0;\eta_z,S_z\rangle \, ,
\end{equation}
where $q_j$ ($q_i$) defines the different locations of the
pseudoparticles (pseudoholes) relatively to
$(18)$ and $N_{ph}^{\alpha(l,l')}$ is the number of
$\alpha (l,l')$ pseudoparticle-pseudohole processes.
When the starting ground state and the final state
of an excitation I have different electronic
numbers, $N_{\sigma}$, it can be decomposed in a
ground-state - ground-state transition and
a pseudoparticle-pseudohole excitation
around the end ground state \cite{Nuno95}.

All the Hamiltonian eigenstates $(15)$ and $(23)$ are states
I. Non-LWS's and non-HWS's are generated by acting raising or
lowering generators $(4)$ and $(5)$ on these states.

In the Hilbert subspace ${\cal H}_I$ spanned by the states I
the Hubbard model can be written in the pseudoparticle basis.
The derivation is as for the $(-1,-1)$ sector and
the Hamiltoniamn $(1)$ reads \cite{Carmelo94}

\begin{eqnarray}
\hat{H} & = & \hat{H}_{SO(4)} + |\mu |\sum_{q}
[1 - \hat{N}_{c(l,l')}(q)] + \mu_0 |H|\sum_{q}
[1 - \hat{N}_{s(l,l')}(q)]\nonumber \\
& = & \hat{H}_{SO(4)} + |\mu |\sum_{q}
[1 - \hat{N}_{c(l,l')}(q)] + \mu_0 |H|[\sum_{q}
\hat{N}_{c(l,l')}(q) - 2\sum_{q} \hat{N}_{s(l,l')}(q)] \, ,
\end{eqnarray}
where the Hamiltonian $\hat{H}_{SO(4)}$ $(2)$ is of the
form

\begin{equation}
\hat{H}_{SO(4)} = \sum_{q}\left[\hat{N}_{c(l,l')}(q)
\{-2t\cos [\hat{K}_{l,l'}(q)]
- U/2\} + U/4\right] \, ,
\end{equation}
$\hat{K}_{l,l'}(q)$ is a rapidity operator \cite{Carmelo94}
whose eigenvalues, $K_{l,l'}(q)$, are studied in Appendix A
[see Eqs. (A3) and (A4)], and $\hat{N}_{c(l,l')}(q)$ is
the $c(l,l')$ pseudomomentum distribution operator.
The operator $\hat{N}_{\alpha (l,l')}(q)$ has the form

\begin{equation}
\hat{N}_{\alpha (l,l')}(q) =
b^{\dag }_{q\alpha (l,l')}b_{q\alpha (l,l')} \, ,
\end{equation}
and we can write the $\alpha (l,l')$ pseudoparticle number
operator as

\begin{equation}
\hat{N}_{\alpha (l,l')} = \sum_{q}\hat{N}_{\alpha (l,l')}(q) \, .
\end{equation}
It follows that the operators $\hat{\eta}_z $ and $\hat{S}_z $
read

\begin{equation}
\hat{\eta}_z = l
\sum_{q}[1 - \hat{N}_{c(l,l')}(q)] \, , \hspace{1cm}
\hat{S}_z = l' \sum_{q}[1 -
\hat{N}_{s(l,l')}(q)] \, .
\end{equation}
This, together with the following
relations valid for all $(l,l')$ sectors,

\begin{equation}
-l \mu = |\mu | \, ;
\hspace{2cm} -l' H = |H| \, ,
\end{equation}
justifies the form $(24)$ of the Hamiltonian $(1)$
in the pseudoparticle basis.

In normal order relatively to the state $(18)$ the expression
of this Hamiltonian involves the normal-ordered operators
\cite{Carmelo94,Carmelo95}

\begin{equation}
:\hat{N}_{q\alpha (l,l')}: =
:b^{\dag }_{q\alpha(l,l')}b_{q\alpha(l,l')}: \, ,
\end{equation}
and has pseudoparticle forward-scattering terms only. (The
expression of the normal-ordered operator
$:b^{\dag }_{q\alpha(l,l')}b_{q\alpha(l,l')}:$ of the right-hand
side (rhs) of Eq. $(30)$ is given in Eq. $(43)$ below.) As in the
case of the $(-1,-1)$ Hamiltonian of Ref. \cite{Carmelo94}, it
has an infinite number of terms which correspond to increasing
scattering orders. To second order we find for each $(l,l')$
sector

\begin{equation}
:\hat{H}: = \sum_{\alpha,q}\epsilon_{\alpha (l,l')}(q)
:\hat{N}_{q\alpha (l,l')}: +
{1\over 2}\sum_{\alpha,q}\sum_{\alpha',q'}
f_{\alpha\alpha'}^{l,l'}(q,q')
:\hat{N}_{q\alpha (l,l')}::\hat{N}_{q'\alpha' (l,l')}:
+ ....\, .
\end{equation}
where the pseudoparticle bands
$\epsilon_{\alpha (l,l')}(q)$ and the $f$ functions
are evaluated as in the case of the $(-1,-1)$ sector
\cite{Carmelo94,Carmelo92b}. Since the colors
$\alpha (l,l')$ and $\alpha '(l,l')$ of two-pseudoparticle
quantities, such as the above $f$ functions and below
phase shifts (see also Appendix A), refer to the same $(l,l')$ numbers,
in order to simplify our notation we call
in this case these colors $\alpha $ and $\alpha '$, respectively.
(The numbers $(l,l')$ need to appear only once in these
functions.) The expressions of the bands are

\begin{equation}
\epsilon_{c(l,l')}(q) = \epsilon_{c(l,l')}^0(q) + [|\mu| -
U/2] - \mu_0 |H| \, ,
\end{equation}
and

\begin{equation}
\epsilon_{s(l,l')}(q) = \epsilon_{s(l,l')}^0(q) + 2\mu_0 |H| \, ,
\end{equation}
where the pseudoparticle bare spectra $\epsilon_{c(l,l')}^0(q)$ and
$\epsilon_{s(l,l')}^0(p)$ are given by

\begin{equation}
\epsilon_{c(l,l')}^0(q) = -2t\cos K^{(0)}_{l,l'}(q) +
2t\int_{-Q_{l,l'}}^{Q_{l,l'}}
dk\widetilde{\Phi }_{cc}^{l,l'}
\left(k,K^{(0)}_{l,l'}(q)\right)\sin k \, ,
\end{equation}
and

\begin{equation}
\epsilon_{s(l,l')}^0(q) = 2t\int_{-Q_{l,l'}}^{Q_{l,l'}}dk
\widetilde{\Phi }_{cs}^{l,l'}
\left(k,S^{(0)}_{l,l'}(q)\right) \sin k \, ,
\end{equation}
respectively. The phase shifts
$\widetilde{\Phi }_{\alpha\alpha'}^{l,l'}$
are defined in Eqs. (A39)-(A42) and the functions
$K^{(0)}_{l,l'}(q)$ and
$S^{(0)}_{l,l'}(q)$ are the solutions of Eqs. (A3)-(A4)
for the reference state $(18)$. The parameters
$Q_{l,l'}$ are given in Eq. (A22). The
functions $K^{(0)}_{l,l'}(q)$ and
$S^{(0)}_{l,l'}(q)$ can be defined in terms of
the phase shifts (A39) and (A40), respectively, as follows

\begin{equation}
K^{(0)}_{l,l'}(q) = q - \int_{-Q_{l,l'}}^{Q_{l,l'}}dk
\widetilde{\Phi }_{cc}^{l,l'}\left(k,K^{(0)}_{l,l'}(q)\right) \, ,
\hspace{1cm} q =
\int_{-Q_{l,l'}}^{Q_{l,l'}}dk
\widetilde{\Phi }_{cs}^{l,l'}\left(k,S^{(0)}_{l,l'}(q)\right) \, .
\end{equation}

The pseudoparticle bands $\epsilon_{\alpha
(-1,-1)}(q)=\epsilon_{\alpha }(q)$ are plotted in Figs. 7
and 8 of Ref. \cite{Carmelo91b}.
In the present $(l,l')$ sectors of parameter space
of symmetry $U(1)\otimes U(1)$
the pseudoparticle energy spectra
$\epsilon_{c(l,l')}(q)$ and $\epsilon_{s(l,l')}(q)$ defined by
Eqs. $(32)$ and $(33)$ vanish at the pseudo-Fermi-points, i.e.,

\begin{equation}
\epsilon_{\alpha (l,l')}(q_{F\alpha (l,l')}^{(\pm )}) = 0 \, .
\end{equation}

The $f$ functions, $f_{\alpha\alpha'}^{l,l'}(q,q')$, of
the rhs of Eq. (31) read

\begin{eqnarray}
f_{\alpha\alpha'}^{l,l'}(q,q') & = & 2\pi
v_{\alpha (l,l')}(q)
\Phi_{\alpha\alpha'}^{l,l'}(q,q')
+ 2\pi v_{\alpha' (l,l')}(q') \Phi_{\alpha'\alpha}^{l,l'}(q',q)
\nonumber \\
& + & \sum_{j=\pm 1} \sum_{\alpha'' =c,s}
2\pi v_{\alpha'' (l,l')}
\Phi_{\alpha''\alpha}^{l,l'}(jq_{F\alpha'' (l,l')},q)
\Phi_{\alpha''\alpha'}^{l,l'}(jq_{F\alpha'' (l,l')},q') \, ,
\end{eqnarray}
where the two-pseudoparticle phase shifts
$\Phi_{\alpha\alpha'}^{l,l'}(q,q')$ are defined by
Eqs. (A18)-(A21) and the pseudoparticle group velocities are given by

\begin{equation}
v_{\alpha (l,l')}(q) = {d\epsilon_{\alpha (l,l')}(q) \over {dq}} \, .
\end{equation}
The ``light'' velocities

\begin{equation}
v_{\alpha (l,l')}\equiv v_{\alpha (l,l')}(q_{F\alpha (l,l')}) \, ,
\end{equation}
play a determining role at the critical point and
appear in the conformal-invariant expressions
\cite{Frahm,Neto,Carmelo94}. (The velocities
$v_{\alpha (-1,-1)}=v_{\alpha}$ are plotted in Fig. 9 of Ref.
\cite{Carmelo91b}.)

Equation $(31)$ is a generalization of the corresponding
$(-1,-1)$ Hamiltonian of Ref. \cite{Carmelo94}
for the three remaining
sectors. We emphasize that in the present $U(1)\otimes U(1)$
sectors of parameter space and at energy scales smaller than
the gaps for the non-LWS's, non-HWS's, and states II
Eqs. $(24)$, $(25)$, and $(31)$ refer to the expression of the full
quantum-liquid Hamiltonian. In the electronic basis this is given by
Eqs. $(1)$ and $(2)$. (In the case of the sector
$(-1,-1)$, the pseudoparticle-operator representation
$(31)$ leads, in a natural way, to the low-energy
spectrum studied in Refs.
\cite{Frahm,Carmelo92,Carmelo92c,Carmelo91b,Carmelo92b}.)

Both the $f$ functions $(38)$ and all the remaining higher order
coefficients have universal forms in terms of the
two-pseudoparticle phase shifts and pseudomomentum derivatives
of the bands and their higher order ``velocities''. The two operators
of the rhs of Eq. $(31)$ are the Hamiltonian terms which are
relevant at low energy \cite{Neto,Carmelo94}.

The perturbative character of the pseudoparticle basis
follows from the fact that, in contrast to the two-electron
forward scattering amplitudes and vertices, the
two-pseudoparticle $f$ functions (given by Eq. (38))
and the corresponding two-pseudoparticle
forward-scattering amplitudes, which were calculated in
Ref. \cite{Carmelo92c} for the $(-1,-1)$ sector, do not diverge
and are finite.

The combination of Eqs. $(32)$, $(33)$, and $(37)$ allows the
derivation of the density and magnetization curves which read

\begin{equation}
\mu = -l [{U \over 2} -
\epsilon_{c(l,l')}^0(q_{Fc (l,l')})\mid_x
- {1 \over 2}\epsilon_{s(l,l')}^0(q_{Fs (l,l')})]\mid_x \, ,
\end{equation}
and

\begin{equation}
H = l' [{\epsilon_{s(l,l')}^0(q_{Fs (l,l')})
\over {2\mu_0 }}]\mid_y  \, ,
\end{equation}
respectively, where, depending on the constraints imposed to
the system, either $x=H$ or $x=m$ and either $y=\mu$ or $y=n$,
respectively.

The excited states I, which in the canonical ensembles of
the four sectors of lowest symmetry are the only gapless states,
never involve pseudoparticles with different $(l,l')$ numbers.
In each of the four $(l,l')$ sectors of symmetry $U(1)\otimes U(1)$
there is a {\it selection rule} which states that out of the eight
pseudoparticle branches {\it only}
pseudoparticle-pseudohole transitions in each of
the two corresponding bands $c(l,l')$ and $s(l,l')$ are
allowed. Therefore, in the canonical ensembles of
the $(l,l')$ sectors the
states I $(15)$, in number given by Eq. $(17)$,
refer to the two $c(l,l')$ and $s(l,l')$ bands only.
This is also true for all states I
associated with higher symmetry sectors \cite{Nuno95}.

\section{THE GROUND STATE IN THE $U(1)\otimes U(1)$ SECTORS}

In this section we use the pseudoparticle basis and the
associated perturbative character of the quantum problem to
justify that in canonical ensembles of the $(l,l')$ sectors
the ground state has the form $(18)$ and
corresponds to filling symmetrically around the origin the
BA quantum numbers.

To show that at given eigenvalues $\eta_z$ and $S_z$
and within all the states I of form $(15)$ and in
total number given by Eq. $(17)$ the Hamiltonian eigenstate
$(18)$ (which is degenerate when its momentum is
finite) is the state of minimal energy, we use the
pseudoparticle basis of ${\cal H}_I$, where the Hamiltonian
$(1)$ and $(2)$ is given by Eqs. $(24)$ and $(25)$,
respectively.

In Appendix B we study the general energy expression
for all states I of ${\cal H}_I$ (with common
$\eta_z$ and $S_z$ eigenvalues), which is given
by Eqs. (B1), (B2), and (B4). In the case of zero-spin
density, $m=0$, the energy $E^0_{SO(4)}+(U/2)[N-N_a/2]$ was studied
and plotted by Shiba \cite{Shiba} [here $E^0_{SO(4)}$
is the energy (B2) and (B4) of the sector
$(-1,-1)$]. In that Appendix we devote
particular attention to the states I of minimal
and maximal energies. The study of the energy
expressions for the different Hamiltonian eigenstates
with common eigenvalues $\eta_z$ and $S_z$
identifies these two states I. The results are:

First, the study of the energy (B1) reveals that the
energies of the states I of form $(15)$ and
with common eigenvalues $\eta_z$ and $S_z$
correspond to a continuous distribution without
energy gaps.

Second, the minimal and maximal energies of that
continuous distribution of energies of states
with common eigenvalues $\eta_z$ and $S_z$ corresponds
to the states I of pseudomomentum distributions
$N_{\alpha (l,l')}(q)=N_{\alpha (l,l')}^0(q)$ and
$N_{\alpha (l,l')}(q)=N_{\alpha (l,l')}^*(q)$, respectively,
where $N_{\alpha (l,l')}^0(q)$ and $N_{\alpha (l,l')}^*(q)$
are given by Eqs. (B3) and (B5), respectively.

Third, since the energies of the states I of form
$(15)$ and with common eigenvalues $\eta_z$ and $S_z$
correspond to a continuous distribution without energy gaps,
in order to show that (B11) and (B12) are the
minimal and maximal values for these energies, respectively,
it is enought to show that the energies $[E-E_0]$ and
$[E-E_*]$ of the sub class of these states that can be
generated from $(18)$ and (B10), respectively, by
changing the distribution occupancies of an arbitrary
small density of pseudoparticles relative to the
distributions (B3) and (B5), are positive
and negative, respectively. Therefore, within
all states $(15)$ in number of $(17)$ it is enough to
evaluate the energies of the Hamiltonian eigenstates which
differ from the reference distributions
(B3) and (B5) by changing the
occupied pseudomomenta by a small density for
the two branches of $c(l,l')$ and $s(l,l')$
pseudoparticles.

Fourth, and in contrast to the electronic basis, we can
use the perturbative character of the pseudoparticle
basis \cite{Carmelo94} to expand the energies
$[E-E_0]/N_a$ and $[E-E_*]/N_a$ of the above
sub class of states in
the densities of excited $\alpha (l,l')$ pseudoparticles,
$n_{ex}^{\alpha (l,l')}$, relative to the reference
distributions (B3) and (B5). Furthermore, we can
write the Hamiltonian $(1)$ with $(2)$ given by $(25)$
in normal order relative to the reference states $(18)$
and (B10) \cite{Carmelo94}. In Sec. II we have presented
the normal-ordered Hamiltonian relative to the state
$(18)$, Eq. $(31)$. The normal-order character of
these Hamiltonians implies that the corresponding
energies are given relative to the states $(18)$ and
(B10), respectively. These Hamiltonians have an infinite
number of terms [see Eq. $(31)$ for the case of the
state $(18)$] and, therefore,
the evaluation of these energies seems to require the
evaluation of an infinite number of energy contributions.
However, the perturbative character of the pseudoparticle
basis makes the problem much easier. This perturbative
character rests on the fact that the evaluation of these energies
up to the $i^{th}$ order in these densities requires
considering {\it only} the corresponding Hamiltonian terms
of scattering orders less than or equal to $i$ \cite{Carmelo94}.
This follows from the linearity of the density of excited
$\alpha (l,l')$ pseudoparticles, which are the elementary ``particles''
of the quantum liquid, in $\delta N_{\alpha (l,l')}(q)=
\langle\eta_z,S_z|:\hat{N}_{\alpha (l,l')}(q):
|\eta_z,S_z \rangle$. Here,
the normal-ordered pseudomomentum distribution
relative to the states $(18)$ and (B10) is
given by

\begin{equation}
:\hat{N}_{\alpha (l,l')}(q): =
b^{\dag }_{q\alpha (l,l')}b_{q\alpha (l,l')}
- N_{\alpha (l,l')}^0(q) \, ,
\end{equation}
and

\begin{equation}
:\hat{N}_{\alpha (l,l')}(q): =
b^{\dag }_{q\alpha (l,l')}b_{q\alpha (l,l')}
- N_{\alpha (l,l')}^*(q) \, ,
\end{equation}
respectively.

Fifth, the perturbative character of the pseudoparticle
basis together with the above analysis implies
that in each canonical ensemble it is enough to consider
the states called, in the case of a $(-1,-1)$ state
$(18)$, (B) and (C)
in Refs. \cite{Neto,Carmelo94}. These states can
also be defined in the general case of the $(l,l')$
state $(18)$ and of the state (B10) and correspond
to a small density of pseudoparticle-pseudohole processes
relative to the reference distributions (B3) and (B5). (The
states (A) of Refs. \cite{Neto,Carmelo94} are associated
with changes, $\Delta N_{\sigma}$, in the number of $\sigma$
electrons.) Moreover, one
needs to evaluate the energies of these states up to second
order in the density of $\alpha (l,l')$ pseudoparticles,
$n_{ex}^{\alpha (l,l')}$,
only. This involves only the one- and two-pseudoparticle
terms of the corresponding normal-ordered Hamiltonian.
[For the case of the state $(18)$ see Eq. $(31)$.]
We find that for the reference states $(18)$ and (B10)
{\it all} such energies are positive and negative, respectively.
(For the case of the $(-1,-1)$ reference state $(18)$
these energies were evaluated in Ref. \cite{Carmelo94},
and equal the energy spectrum studied in Ref. \cite{Frahm}
when the number of both up-spin and down-spin electrons
is kept constant.) This confirms that in each canonical
ensemble of a $(l,l')$ sector $(18)$ and (B10) are, within
all states I of form $(15)$, with common eigenvalues
$\eta_z $ and $S_z $, and in number of $(17)$, the Hamiltonian
eigenstates of minimal and maximal energies, respectively.
This implies that the energies $E$, Eq. (B1), of all remaining
states I of that ensemble are such that

\begin{equation}
E_0<E<E_* \, ,
\end{equation}
where $E_0$ and $E_*$ are the energies (B11) and (B12),
respectively. In addition, for each canonical ensemble of
eigenvalues $\eta_z$ and $S_z$, $[E_*-E_0]$ gives the energy
width of the continuous distribution of energies
corresponding to the whole set of $(l,l')$
states I of form $(15)$.

We emphasize that only writing the Hamiltonian in
the pseudoparticle-operator basis introduces the
perturbative character of the quantum problem which
simplified the above analysis. Here we have compared
the energies of the states I only. Therefore, a complete proof
of $(18)$ being the Hamiltonian eigenstate of minimal energy
in canonical ensembles of the present lowest-symmetry sectors
of parameter space requires the evaluation of the energy gaps
of the non-LWS's and non-HWS's and of the states II relative
to that state, which have to be finite.
While the gaps of the non-LWS's and non-HWS's are evaluated
in Sec IV, the gaps of the states II are calculated elsewhere
\cite{Nuno94}.

Finally, the study of the spectrum of the states II of
the sectors of higher symmetry $SO(4)$, $SU(2)\otimes U(1)$,
and $U(1)\otimes SU(2)$ \cite{Nuno95,Nuno94} reveals
that in canonical ensembles of these sectors the ground
state is also a state I of form $(18)$, as we discuss
in Ref. \cite{Nuno95}. [When $\eta_z =0$ and (or )
$S_z =0$, the $\alpha (\pm 1,l')$ and (or )
$\alpha (l,\pm 1)$ pseudoparticles correspond to alternative
representations of the same ground state.] Therefore, in
{\it all} canonical ensembles of the quantum problem and
in the pseudoparticle basis the ground state is a simple
Slater determinant of pseudoparticle levels of the universal
form given by Eq. $(18)$.

\section{ENERGY GAPS OF THE NON-LWS's AND NON-HWS's}

In Sec. III we have used the  pseudoparticle basis to show that
among all $(l,l')$ states I with common
eigenvalues $\eta_z$ and $S_z$, in number of $(17)$ and of the
form $(15)$, the Hamiltonian eigenstate $(18)$ has minimal
energy. Therefore, if in the corresponding canonical ensemble
both all the non-LWS's and non-HWS's and all the states II
(with common $\eta_z$ and $S_z$ eigenvalues) have a gap
relative to the state $(18)$, this state
is the ground state. In this section we show that all
such non-LWS's and non-HWS's have an energy gap
and calculate the smallest of these gaps.

Let us consider the Hamiltonian $\hat{H}_{SO(4)}$, Eq. $(2)$.
This represents the Hubbard chain $(1)$ at zero magnetic field and
chemical potential and commutes with the six generators of
Eqs. $(3)-(5)$ and, therefore, has $SO(4)$ symmetry
\cite{Korepin}, as discussed in Sec. II.

Let $|\eta ,S; \eta_z , S_z\rangle $ be an
arbitrary Hamiltonian eigenstate belonging the
family of states with fixed eta spin and spin,
$\eta$ and $S$, respectively. In
$|\eta ,S; \eta_z , S_z\rangle $ $\eta_z$ and $S_z$
are the eta-spin and spin projections, respectively,
of this particular state. The whole family of states
with the same values of eta spin $\eta$ and spin $S$
but different eta-spin and spin projections
can be generated from one of the four [LWS,LWS],
[LWS,HWS], [HWS,LWS], and [HWS,HWS] with
these values of eta spin and spin. The [LWS,LWS] starting state,
for example, is the state $|\eta ,S; -\eta ,
-S\rangle $ which belongs to the sector $(-1,-1)$.
The remaing three choices for starting states are
$|\eta ,S; -\eta , S\rangle $, $|\eta ,S; \eta ,
-S\rangle $, and $|\eta ,S; \eta , S\rangle $, respectively.
Depending on the LWS or HWS character
of the starting state, the family of states
is generated by acting operators $\hat{\eta}$
and $\hat{S}$ or ${\hat{\eta}}^{\dag }$ and
${\hat{S}}^{\dag }$ of Eqs. $(4)$ and $(5)$
onto that state.

In Appendix C we consider fixed values of the
chemical potential $\mu$  and magnetic field
$H$ such that $\mu\neq 0$ and $H\neq 0$.
The signs of the chemical potential and magnetic field
fix the signs of $\eta_z$ and $S_z$ and choose the
particular $(l,l')$ sector. According to Eq. $(29)$, in the sectors
$(-1,-1)$, $(-1,1)$, $(1,-1)$, and $(1,1)$ the
chemical potential $\mu$ and magnetic field $H$ are
such that $\mu >0$ and $H>0$, $\mu >0$ and $H<0$, $\mu <0$
and $H>0$, and $\mu <0$ and $H<0$, respectively.
In that Appendix we show that at fixed and finite values
of the chemical potential and magnetic field
the lowest energy Hamiltonian eigenstate of
a family of states $|\eta ,S; \eta_z , S_z\rangle $
with common values of $\eta$ and $S$ but different
values of $\eta_z$ and $S_z$ is the
[LWS,LWS], [LWS,HWS], [HWS,LWS], or
[HWS,HWS] corresponding to the $(l,l')$
sector choosed by the signs of the chemical
potential and magnetic field. We have also
calculated the smallest energy gaps relative
to that state, which are given in Eq. (C11).

The main goal of this section is, however, to show that
within all states with different $\eta$ and $S$ values but
the same eigenvalues $\eta_z$ and $S_z$, i.e. of
states belonging the same canonical ensemble,
the ground state $(18)$ has minimal energy.

Following the results of Sec. III, within all states I
with the same eigenvalues $\eta_z$ and $S_z$ the state of
minimal energy has the form $(18)$. Therefore,
we can restrict our considerations to the
set of non-LWS's and non-HWS's, $|\eta ,S; \eta_z , S_z\rangle $,
whose starting LWS's and (or) HWS's, $|\eta ,S;\pm\eta ,\pm S\rangle $,
are ``ground states'' of the form $(18)$.
Other non-LWS's and non-HWS's with the same
$\eta_z$ and $S_z$ eigenvalues are of higher energy.

The non-LWS's and non-HWS's belong the same canonical
ensemble and, therefore, have common ${\eta }_z$ and
$S_z$ eigenvalues. We emphasize that their starting states,
$|\eta ,S;\pm\eta ,\pm S\rangle $, do not belong to that
canonical ensemble. In the corresponding $(l,l')=(sgn (\eta_z)1,sgn (S_z)1)$
sector of symmetry $U(1)\otimes U(1)$, the energy of the ground
state $(18)$ of eigenvalues ${\eta }_z$ and
$S_z$ can be written as

\begin{equation}
E^0 = E^0_{SO(4)}(|\eta_z| ,|S_z|) +
2\mu {\eta }_z + 2\mu_0 HS_z \, ,
\end{equation}
where $E^0_{SO(4)}(\eta ,S)$ is the corresponding eigenenergy
relative to the $SO(4)$-Hamiltonian $(2)$. Note that
$E^0_{SO(4)}(\eta ,S)$ is nothing but the term $E^0_{SO(4)}$
of the ground-state energy (B11). Its $\eta$ and $S$
dependence can be obtained from that expression by replacing
the density and spin-density dependences by $\eta$ and $S$ dependences,
respectively. The energy $E^0_{SO(4)}(\eta ,S)$ reads

\begin{equation}
E^0_{SO(4)}(\eta ,S) = {N_a\over {2\pi}}
\int_{-\pi}^{\pi}dk2\pi\rho^0_{c(l,l')}(k)
\left[\Theta (Q_{l,l'}-|k|)\{-2t\cos k
- U/2\} + U/4\right] \, ,
\end{equation}
where the function $\rho^0_{c(l,l')}(k)$ is associated
with the function $\rho^0_{s(l,l')}(v)$ through the
integral equations (A33) and (A34) for the particular
case of the ground-state distributions (B13) with
$Q_{l,l'}$ defined in Eq. (A22). These two
coupled integral equations have a unique solution which
defines the functions $\rho^0_{c(l,l')}(k)$ and
$\rho^0_{s(l,l')}(v)$. For fixed $U$ the dependence of
the energy $(47)$ and functions $\rho^0_{c(l,l')}(k)$ and
$\rho^0_{s(l,l')}(v)$ on the $\eta$ and $S$ values is defined
by the following normalization equations

\begin{equation}
{N_a\over {4\pi}}\int_{-\pi}^{\pi}
dk2\pi\rho^0_{c(l,l')}(k)[1 - \Theta (Q_{l,l'}-|k|)] = \eta \, ,
\end{equation}
and

\begin{equation}
{N_a\over {4\pi}}\int_{-\infty}^{\infty}
dv2\pi\rho^0_{s(l,l')}(v)[1 - \Theta (B_{l,l'}/u-|v|)] = S \, .
\end{equation}
These equations also define the dependence on
$\eta $ and $S$ of the parameters $Q_{l,l'}$ and $B_{l,l'}$
of Eq. (A22).
The use of Eqs. $(13)$ and $(14)$ reveals that $(48)$ and
$(49)$ are equivalent to the normalization conditions
(A36) and (A38) for the particular case of the
ground-state distributions (B13). The above integral
equations fully define the energy $E^0_{SO(4)}(\eta ,S)$
and can be solved numerically. (A closed-form analytical
solution is available for $E^0_{SO(4)}(0,0)$ \cite{Lieb}.)

On the other hand, the energy $E(\eta , S)$ of a non-LWS
and non-HWS, $|\eta ,S; \eta_z , S_z\rangle$, with
the same values of ${\eta }_z$ and $S_z$ (but
${\eta }_z\neq \pm\eta$ and (or) $S_z\neq \pm S$)
is, following Eq. (C2), given by

\begin{equation}
E(\eta , S) = E^0_{SO(4)}(\eta ,S) +
2\mu {\eta }_z + 2\mu_0 HS_z \, ,
\end{equation}
where the energy $E^0_{SO(4)}(\eta ,S)$, Eq. $(47)$,
refers to the corresponding starting states
$|\eta ,S;\pm\eta ,\pm S\rangle $. Following our
choice, these states are also ground states
of form $(18)$, but such that $\eta\neq |\eta_z|$
and (or ) $S\neq |S_z|$.

We want to show that the energy gap

\begin{equation}
E(\eta , S) - E^0 = E^0_{SO(4)}(\eta ,S) -
E^0_{SO(4)}(|\eta_z| ,|S_z|) \, ,
\end{equation}
where $E^0$, $E(\eta , S)$, and $E^0_{SO(4)}(\eta ,S)$
(and $E^0_{SO(4)}(|\eta_z| ,|S_z|)$) are given by
Eqs. $(46)$, $(50)$, and $(47)$, respectively,
is positive for ${\eta }_z\neq \pm\eta$ and (or) $S_z\neq
\pm S$. The energy $E^0_{SO(4)}(\eta ,S)$, Eq. $(47)$, is a monotonous
increasing function of both $\eta$ and $S$, its minimum value
being $E^0_{SO(4)}(0 ,0)$. This refers to the ``absolute''
$SO(4)$ ground state \cite{Lieb,Nuno95}. Since $\eta >|\eta_z|$ and
(or) $S >|S_z|$ [if $S =|S_z|$ (or $\eta =|\eta_z|$) we have
that $\eta >|\eta_z|$ (or $S >|S_z|$)], it follows that
the gap $(51)$ is always positive. Its smallest values correspond to
the choices (a) $\eta =|\eta_z|+1$ and $S =|S_z|$;
and (b) $\eta =|\eta_z|$ and $S =|S_z|+1$. Let us
evaluate the corresponding gaps

\begin{equation}
\Delta_a = E^0_{SO(4)}(|\eta_z|+1,|S_z|) -
E^0_{SO(4)}(|\eta_z| ,|S_z|) \, ,
\end{equation}
and

\begin{equation}
\Delta_b = E^0_{SO(4)}(|\eta_z|,|S_z|+1) -
E^0_{SO(4)}(|\eta_z| ,|S_z|) \, ,
\end{equation}
respectively. The gaps $(52)$ and $(53)$ give the excitation
energy of the states $||\eta_z|+1,|S_z|;\eta_z,S_z\rangle$
and $||\eta_z|,|S_z|+1;\eta_z,S_z\rangle$, respectively,
relative to the LWS's and (or) HWS's, $||\eta_z|,|S_z|;
\eta_z,S_z\rangle$, of the form $(18)$.
To evaluate these gaps we use the fact
that in the thermodynamic limit the excitation
energy of the LWS's and (or) HWS's
$||\eta_z|+1,|S_z|;\eta_z+sgn (\eta_z)1,S_z\rangle$
and $||\eta_z|,|S_z|+1;\eta_z,S_z+sgn (S_z)1\rangle$
[also of form $(18)$] relative to the LWS and (or) HWS
$||\eta_z|,|S_z|; \eta_z,S_z\rangle$ is zero. This excitation
is the spin-flip ground-state -- ground-state
transition (i)-(iv) studied in Ref. \cite{Nuno95}.
Its excitation energy is of order $1/N_a$ and
vanishes in the present thermodynamic limit.
(This is a condition for the continuous character of the
magnetization curve defined by Eq. $(42)$.) This implies that

\begin{equation}
E^0_{SO(4)}(|\eta_z| ,|S_z|) +
2\mu {\eta }_z + 2\mu_0 HS_z =
E^0_{SO(4)}(|\eta_z|+1,|S_z|) +
2\mu [{\eta }_z+sgn (\eta_z)1] + 2\mu_0 HS_z \, ,
\end{equation}
and

\begin{equation}
E^0_{SO(4)}(|\eta_z| ,|S_z|) +
2\mu {\eta }_z + 2\mu_0 HS_z =
E^0_{SO(4)}(|\eta_z|,|S_z|+1) +
2\mu {\eta }_z + 2\mu_0 H[S_z+sgn (S_z)1] \, .
\end{equation}
Taking into account the relation between the signs
of $\mu$ (and $H$) and of $\eta_z$ (and $S_z$)
[see Eq. $(29)$], and combining Eqs. $(52)$ and $(53)$ with Eqs.
$(54)$ and $(55)$, we finally arrive to:

\begin{equation}
\Delta_a = 2|\mu| \, ;\hspace{2cm} \Delta_b =
2\mu_0 |H| \, .
\end{equation}
These are the smallest gaps of non-LWS's and non-HWS's
belonging the same canonical ensemble (i.e. having the
same eigenvalues $\eta_z$ and $S_z$) relative to the corresponding
ground state $(18)$. Note that both (or one) of these
gaps vanish(es) in the $SO(4)$ sector [or $SU(2)\otimes U(1)$
and $U(1)\otimes SU(2)$ sectors] of parameter space.
On the other hand, all Hamiltonian eigenstates with
$\eta =0$ and (or ) $S=0$ are both LWS's and HWS's of the
corresponding algebras. This implies that in the canonical
ensembles with $\eta =0$ and (or ) $S=0$ there are no
non-LWS's and non-HWS's singlets of the eta-spin and (or ) spin
algebras.

\section{CONCLUDING REMARKS}

In this paper we have used the pseudoparticle-operator basis
and perturbation theory introduced in Refs.
\cite{Neto,Carmelo94,Carmelo95}
to derive and study the ground states associated with all
canonical ensembles belonging to the four $U(1)\otimes U(1)$
sectors of parameter space of the Hubbard chain in the
presence of a magnetic field and chemical potential.
Our results confirm the important role played
by the pseudoparticle algebra
in the low-energy physics of integrable quantum liquids:
following the present study we find in Ref. \cite{Nuno95}
that the usual half-filling and zero-magnetic-field
holons, antiholons, and spinons \cite{Essler} correspond to a limiting
case of the general pseudoparticle representation.

The simple form obtained for the ground-state, expression $(18)$,
has a deep physical meaning. It confirms \cite{Neto,Carmelo94}
and generalizes the fact that in the pseudoparticle basis
the ground state of the many-electron quantum problem is a
``non-interacting'' pseudoparticle ground state of simple
Slater-determinant form. This also holds true for canonical
ensembles belonging to sectors of higher symmetry
\cite{Nuno95,Nuno94} and, therefore, in the pseudoparticle
basis the ground state of canonical ensembles
of all symmetries are {\it always} states I of
that simple form.

We have evaluated the energy gaps relative to the
ground state of the non-LWS's and non-HWS's with
common $\eta_z $ and $S_z $ eigenvalues. A
complete proof of our ground-state expressions
requires the calculation of the energy gaps
of the states II with the same $\eta_z $ and $S_z $
eigenvalues \cite{Nuno94}.

A more general Landau-liquid theory for the
sectors $U(1)\otimes U(1)$ including
the states II can be constructed. These states can also
be described in terms of pseudoparticles.
However, in the sectors of lowest symmetry these
requires, in addition to the pseudoparticles
studied in this paper, new branches of
``heavy'' pseudoparticles \cite{Nuno94}.

The eight branches of $\alpha (l,l')$ pseudoparticles
introduced in this paper have a deep physical meaning.
This is shown in Ref. \cite{Nuno95} where we
relate the symmetry transformations of
the set of pseudoparticles used in each canonical ensemble
to construct the corresponding ground state
to the symmetry of that sector. In that reference we
find that the pseudoparticles of this set always transform
in the representation of the corresponding
group of symmetry.

Although the pseudoparticles
associated with the states I are the transport carriers
at low energy \cite{Carmelo92c} and couple to external
potentials \cite{Campbell}, they refer to purely
non-dissipative excitations, {\it i.e.} the Hamiltonian
{\it commutes} with the currents in the subspace
spanned by the states I \cite{Carmelo92c}. Therefore,
these pseudoparticle currents give
rise {\it only} to the coherent part of the conductivity
spectra, i.e. to the Drude peaks \cite{Carmelo92c,Neto,Carmelo94}.
The finite-frequency part is associated with transitions
involving the ``heavy'' pseudoparticles which
are also needed, in the sectors of symmetry $U(1)\otimes U(1)$,
to describe the states II \cite{Nuno94}. (In the Hilbert
subspace spanned by those excitations, the Hamiltonian
{\it does not} commute with the current operators.)

As in the case of Landau's Fermi liquid theory
\cite{Pines,Baym}, the pseudoparticle perturbation theory
uses as reference state the exact ground state of the quantum
problem \cite{Neto,Carmelo94}. Also in the construction of
the above generalized Landau-liquid theory referring to the
Hilbert space spanned by both the states I and states II
\cite{Nuno94}, the ground state which we have investigated
and studied in the present paper plays a crucial role.

\nonum
\section{ACKNOWLEDGMENTS}

This work was supported principally by the C.S.I.C. [Spain] and
the Faculty of Sciences of the University of Lisbon. We thank
D. K. Campbell, F. Guinea, and A. H. Castro Neto for stimulating
discussions. N. M. R. P. is grateful for the support of A.
A. Barroso and Faculty of Sciences of the University of Lisbon
and for the hospitality of the C.S.I.C. [Madrid].

\vfill
\eject
\appendix{BETHE-ANSATZ SOLUTION OF THE FOUR $U(1)\otimes U(1)$ SECTORS}

The BA solution associated with the Hamiltonian eigenstates
I of the four $(l,l')$ sectors is very similar to the
solution of the $(-1,-1)$ sector studied in
Refs. \cite{Lieb,Frahm,Carmelo94}.

In this Appendix we present the BA equations for the
states I of the $(l,l')$ sectors of symmetry
$U(1)\otimes U(1)$ and introduce the two-pseudoparticle
phase shifts \cite{Carmelo92c,Carmelo94,Carmelo92b}
and other quantities needed in the
expressions presented in this paper.

For each choice of $N_{c(l,l')}$ and $N_{s(l,l')}$
occupied pseudomomenta values $q_j$, Eq. $(12)$, of the
$c(l,l')$ and $s(l,l')$ pseudoparticles, respectively,
(see the expressions of these numbers in terms of
electronic numbers in Table 1) which
describes one state I, Eq. $(15)$, there is a set
of $N_{c(l,l')}$ real rapidity values,  $k^{l,l'}_j$,
and other $N_{s(l,l')}$ real rapidity values, $v^{l,l'}_j$,
which are the solution of the following
$N_{c(l,l')}+N_{s(l,l')}$ algebraic equations

\begin{equation}
k^{l,l'}_j = q_j + {2\over N_a}\sum_{j'=1}^{N_{s(l,l')}}
\tan^{-1}\Bigl(v^{l,l'}_{j'} -
\left({1/u}\right)\sin k^{l,l'}_j\Bigr) \, ,
\hspace{1cm} j=1,...,N_{c(l,l')} \, ,
\end{equation}
and

\begin{eqnarray}
q_j & = & {2\over N_a}\sum_{j'=1}^{N_{c(l,l')}}
{\tan^{-1}\Bigl(v^{l,l'}_j - \left({1/u}
\right)\sin k^{l,l'}_{j'} \Bigr)} \nonumber \\
& - & {2\over N_a}\sum_{j'=1}^{N_{s(l,l')}}
\tan^{-1}\Bigl({1\over 2}\left(v^{l,l'}_j -
v^{l,l'}_{j'} \right)\Bigr) \, ,
\hspace{1cm} j=1,...,N_{s(l,l')} \, .
\end{eqnarray}

In the thermodynamic limit ($N_{\alpha (l,l')},N_a\rightarrow\infty$
with $n_{\alpha (l,l')}=N_{\alpha (l,l')}/N_a$ finite)
the rapidity values
$k^{l,l'}_j$ and $v^{l,l'}_j$ give rise to
rapidity functions $K_{l,l'}(q)$ and $S_{l,l'}(q')$,
respectively, which are eigenvalues of the
corresponding rapidity operators \cite{Carmelo94}.
The set of algebraic equations (A1)-(A2) lead to
the following two coupled integral equations

\begin{equation}
K_{l,l'}(q)=q+{1\over\pi}\int_{q^{(-)}_{s(l,l')}}^{q^{(+)}_{s(l,l')}}dq'
N_{s(l,l')}(q')\tan^{-1}\Bigl(S_{l,l'}(q')-
\left({1/u}\right)\sin K_{l,l'}(q)\Bigr) \, ,
\end{equation}
and

\begin{eqnarray}
q & = & {1\over\pi}\int_{q^{(-)}_{c(l,l')}}^{q^{(+)}_{c(l,l')}}
{dq'N_{c(l,l')}(q')\tan^{-1}\Bigl(S_{l,l'}(q)-\left({1/u}
\right)\sin K_{l,l'}(q')\Bigr)} \nonumber \\
& - & {1\over\pi}\int_{q^{(-)}_{s(l,l')}}^{q^{(+)}_{s(l,l')}}dq'
N_{s(l,l')}(q')\tan^{-1}\Bigl({1\over 2}\left(S_{l,l'}(q)-
S_{l,l'}(q')\right)\Bigr) \, ,
\end{eqnarray}
respectively, where the limits of the pseudo-Brillouin zones,
$q^{(\pm)}_{\alpha (l,l')}$, are given by Eqs. $(19)-(21)$
with $N_{\alpha (l,l')}$ replaced by $N_{\alpha (l,l')}^*$
and the pseudomomentum distributions, $N_{c(l,l')}(q)$, are
the eigenvalues (and also expectation values) of the operators
$(26)$ relatively to the states I of form $(15)$, ie

\begin{equation}
N_{\alpha (l,l')}(q) = \langle
\eta_z,S_z|\hat{N}_{\alpha (l,l')}(q)
|\eta_z,S_z\rangle
\, .
\end{equation}
The pseudomomentum distribution
$N_{\alpha (l,l')}(q)$ (A5) is $1$ and $0$ for occupied and
nonoccupied pseudomomenta, respectively, of the states $(15)$.
Therefore, the distributions (A5) fully define these
Hamiltonian eigenstates I [see Eq. $(27)$]. We have that

\begin{equation}
{N_a\over 2\pi}
\int_{q_{\alpha (l,l')}^{(-)}}^{q_{\alpha (l,l')}^{(+)}}
dq = N_{\alpha (l,l')}^* \, , \hspace{1cm}
{N_a\over 2\pi}
\int_{q_{\alpha (l,l')}^{(-)}}^{q_{\alpha (l,l')}^{(+)}}
dq N_{\alpha (l,l')}(q) = N_{\alpha (l,l')} \, ,
\end{equation}
where $N_{\alpha (l,l')}^*$ and $N_{\alpha (l,l')}$ are
the number of available $\alpha (l,l')$ pseudomomentum
values , $q_j$ [see Eq. $(12)$], and numbers of $\alpha (l,l')$
pseudoparticles, respectively, given in Table 1.

For each Hamiltonian eigenstate $(15)$ there
is one, and only one, pair of rapidity eigenvalues
$K_{l,l'}(q)$ and $S_{l,l'}(q)$. These are functionals of the
pseudomomentum distributions. The solution of
Eqs. (A3) and (A4) provides these rapidity functionals of the
pseudomomentum distributions $N_{\alpha (l,l')}(q)$.

It is easier to express the rapidity functions in
terms of the eigenvalues of the normal-ordered
operators $(30)$ and $(43)$, which define the pseudomomentum
deviations. The rapidity functions can then
be expanded in these deviations as
\cite{Carmelo94,Carmelo92b}

\begin{equation}
K_{l,l'}(q) = K^{(0)}_{l,l'}(q) + K^{(1)}_{l,l'}(q) +
K^{(2)}_{l,l'}(q) + ...
\end{equation}
and

\begin{equation}
S_{l,l'}(q) = S^{(0)}_{l,l'}(q) + S^{(1)}_{l,l'}(q) +
S^{(2)}_{l,l'}(q) + ...
\end{equation}
where $K^{(j)}_{l,l'}(q)$ and $S^{(j)}_{l,l'}(p)$ are the
$j$th-order terms.
Equations (A3) and (A4) allow the systematic evaluation
order by order of all terms of the expansions (A7) and (A8).
As shown in Ref. \cite{Carmelo94} for the case of
the $(-1,-1)$ sector, this deviation expansion
corresponds to a operator expansion in the pseudoparticle
scattering order. The possibility of such expansion
follows from the perturbative character of the
pseudoparticle operator basis \cite{Neto,Carmelo94,Carmelo92b}.

Here we are interessed in the first-order terms of
(A7) and (A8) which involve the two-pseudoparticle phase shifts
\cite{Carmelo92c,Carmelo94,Carmelo92b}.
By using a recursion procedure, we find
that the rapidities (A7) and (A8) may be simply written as

\begin{equation}
K_{l,l'}(q)=K^{(0)}_{l,l'}({\cal Q}_{l,l'}(q)) \, , \hspace{2cm}
S_{l,l'}(q)=S^{(0)}_{l,l'}({\cal P}_{l,l'}(q)) \, ,
\end{equation}
where $K^{(0)}_{l,l'}(q)$ and $S^{(0)}_{l,l'}(q)$ are the
solutions that correspond to the choice of
distribution (B3) of Appendix B (in this paper
we want to confirm that this choice defines the
ground state) and ${\cal Q}_{l,l'}(q)$
and ${\cal P}_{l,l'}(q)$ are functionals of the form

\begin{equation}
{\cal Q}_{l,l'}(q) = q + {\cal Q}^{(1)}_{l,l'}(q) +
{\cal Q}^{(2)}_{l,l'}(q) + ...\, ,
\end{equation}

\begin{equation}
{\cal P}_{l,l'}(q) = q + {\cal P}^{(1)}_{l,l'}(q) +
{\cal P}^{(2)}_{l,l'}(q) + ...\, ,
\end{equation}
which can be obtained by solving Eqs. (A3) and (A4)
order by order. The
results (A9)-(A11) imply that the first-order terms of
the rhs of Eqs. (A7) and (A8) may be written as

\begin{equation}
K^{(1)}_{l,l'}(q) = {dK^{(0)}_{l,l'}(q)\over {dq}}
{\cal Q}^{(1)}_{l,l'}(q) \, ,
\end{equation}
and

\begin{equation}
S^{(1)}_{l,l'}(q) = {dS^{(0)}_{l,l'}(q)\over {dq}}
{\cal P}^{(1)}_{l,l'}(q) \, ,
\end{equation}
respectively. We note that the functions
$dK^{(0)}_{l,l'}(q)/dq$ and $dS^{(0)}_{l,l'}(q)/dq$
obey the equations

\begin{equation}
{dK^{(0)}_{l,l'}(q)\over dq} =
{1\over 2\pi\rho_{c(l,l')}^0\left(K^{(0)}_{l,l'}(q)\right)} \, ,
\end{equation}
and

\begin{equation}
{dS^{(0)}_{l,l'}(q)\over dq} =
{1\over 2\pi\rho_{s(l,l')}^0\left(S^{(0)}_{l,l'}(q)\right)}\, ,
\end{equation}
respectively, where the functions $2\pi\rho_{c(l,l')}^0(k)$
and $2\pi\rho_{s(l,l')}^0(v)$ are the ``ground-state''
solutions of Eqs. (A33) and (A34) below. [The
functions $2\pi\rho_{c(-1,-1)}^0(k)$ and
$2\pi\rho_{s(-1,-1)}^0(v)$ (with $\rho_{s(-1,-1)}^0(v)
=u\sigma (uv)$, where $v=\Lambda/u$) are the usual ground-state
distributions of Lieb and Wu \cite{Lieb}.]

Solving Eqs. (A3) and (A4) to first order leads to

\begin{equation}
{\cal Q}^{(1)} (q) = \sum_{\alpha }
\int_{q^{(-)}_{\alpha (l,l')}}^{q^{(+)}_{\alpha (l,l')}}
dq'\delta N_{l,l'}(q')\Phi_{c\alpha }^{l,l'}(q,q') \, ,
\end{equation}

\begin{equation}
{\cal P}^{(1)} (q) = \sum_{\alpha }
\int_{q^{(-)}_{\alpha (l,l')}}^{q^{(+)}_{\alpha (l,l')}}
dq'\delta N_{l,l'}(q')\Phi_{s\alpha }^{l,l'}(q,q') \, ,
\end{equation}
where $\delta N_{\alpha (l,l')}(q) = \langle
\eta_z,S_z|:\hat{N}_{\alpha (l,l')}(q):
|\eta_z,S_z\rangle$.
The four two-pseudoparticle phase shifts
$\Phi_{\alpha\alpha '}^{l,l'}(q,q')$
can be written as

\begin{equation}
\Phi_{cc}^{l,l'}(q,q') =
\bar{\Phi }_{cc}^{l,l'}\left({\sin K^{(0)}_{l,l'}(q)\over u},
{\sin K^{(0)}_{l,l'}(q')\over u}\right) \, ,
\end{equation}

\begin{equation}
\Phi_{cs}^{l,l'}(q,q') =
\bar{\Phi }_{cs}^{l,l'}\left({\sin K^{(0)}_{l,l'}(q)\over u},
S^{(0)}_{l,l'}(q')\right) \, ,
\end{equation}

\begin{equation}
\Phi_{sc}^{l,l'}(q,q') =
\bar{\Phi }_{sc}^{l,l'}\left(S^{(0)}_{l,l'}(q),
{\sin K^{(0)}_{l,l'}(q')\over u}\right) \, ,
\end{equation}

\begin{equation}
\Phi_{ss}^{l,l'}(q,q') =
\bar{\Phi }_{ss}^{l,l'}\left(S^{(0)}_{l,l'}(q),
S^{(0)}_{l,l'}(q')\right) \, .
\end{equation}

Introducing the parameters

\begin{equation}
Q_{l,l'} = K^{(0)}_{l,l'}(q_{Fc(l,l')}) \, , \hspace{1cm}
B_{l,l'}/u = S^{(0)}_{l,l'}(q_{Fs(l,l')}) \, ,
\end{equation}
and

\begin{equation}
x^{0}_{l,l'} = {\sin Q_{l,l'}\over u} \, , \hspace{1cm}
y^{0}_{l,l'} = B_{l,l'}/u \, ,
\end{equation}
($Q_{-1,-1}$ and $B_{-1,-1}$ are the usual cutoff
parameters of the ground-state Lieb-Wu equations
\cite{Lieb}) we find that the phase shifts
$\bar{\Phi }_{\alpha\alpha '}^{l,l'}$
obey the following integral equations:

\begin{equation}
\bar{\Phi }_{cc}^{l,l'}\left(x,x'\right) =
{1\over{\pi}}\int_{-y^{0}_{l,l'}}^{y^{0}_{l,l'}}
dy''{\bar{\Phi }_{sc}^{l,l'}\left(y'',x'\right)
\over {1+(x-y'')^2}} \, ,
\end{equation}

\begin{equation}
\bar{\Phi }_{cs}^{l,l'}\left(x,y'\right) =
-{1\over{\pi}}\tan^{-1}(x-y') + {1\over{\pi}}
\int_{-y^{0}_{l,l'}}^{y^{0}_{l,l'}}
dy''{\bar{\Phi }_{ss}^{l,l'}\left(y'',y'\right)
\over {1+(x-y'')^2}} \, ,
\end{equation}

\begin{equation}
\bar{\Phi }_{sc}^{l,l'}\left(y,x'\right) =
-{1\over{\pi}}\tan^{-1}(y-x') + \int_{-y^{0}_{l,l'}}^{y^{0}_{l,l'}}
dy''G(y,y''){\bar{\Phi }}_{sc}^{l,l'}\left(y'',x'\right) \, ,
\end{equation}

\begin{eqnarray}
\bar{\Phi }_{ss}^{l,l'}\left(y,y'\right) & =
& {1\over{\pi}}\tan^{-1}({y-y'\over{2}}) -
{1\over{\pi^2}}\int_{-x^{0}_{l,l'}}^{x^{0}_{l,l'}}dx''{\tan^{-1}
(x''-y')\over{1+(y-x'')^2}}
\nonumber \\
& + & \int_{-y^{0}_{l,l'}}^{y^{0}_{l,l'}}
dy''G(y,y''){\bar{\Phi }}_{ss}^{l,l'}\left(y'',y'\right) \, .
\end{eqnarray}
The kernel $G(y,y')$ reads \cite{Carmelo92b}

\begin{equation}
G(y,y') = - {1\over{2\pi}}\left[{1\over{1+((y-y')/2)^2}}\right]
\left[1 - {1\over 2}
\left(t(y)+t(y')+{{l(y)-l(y')}\over{y-y'}}\right)\right] \, ,
\end{equation}
where

\begin{equation}
t(y) = {1\over{\pi}}\sum_{j=\pm 1}(j)\tan^{-1}(y +
jx^{0}_{l,l'}) \, ,
\end{equation}
and

\begin{equation}
l(y) = {1\over{\pi}}\sum_{j=\pm 1}(j)\ln (1+(y +
jx^{0}_{l,l'})^2) \, .
\end{equation}

It is useful to introduce an alternative representation
for the rapidity functions $K_{l,l'}(q)$ and
$S_{l,l'}(q)$ of Eqs. (A3) and (A4) in terms of
distributions $\rho_{c(l,l')}(k)$ and $\rho_{s(l,l')}(v)$.
This leads to new equations which are equivalent to
the latter equations. This second representation
is less appropriate for the pseudoparticle operator
basis but was, historically, the most widely used
in BA problems \cite{Lieb,Frahm,Shiba}. The reason is that it leads to
integral equations which, in some limits, are of mathematical
standard type. However, and as we discuss below,
the representation associated with the Eqs. (A3) and
(A4) has a clearer physical connection to the
BA operator algebra.

Let us introduce the function $\rho_{c(l,l')}(k)$
such that

\begin{equation}
2\pi\rho_{c (l,l')}(K_{l,l'}(q)) =
1/[{d K_{l,l'}(q)\over {dq}}] \, .
\end{equation}
This function is related to the distribution
$\rho_{s(l,l')}(v)$ which is defined as

\begin{equation}
2\pi\rho_{s(l,l')}(S_{l,l'}(q)) =
1/[{d S_{l,l'}(q)\over {dq}}] \, .
\end{equation}
Equations (A31) and (A32) define the transformation
$q\rightarrow k$ for $\alpha (l,l') =c(l,l')$ and
$q\rightarrow v$ for $\alpha (l,l')=s(l,l')$, where the
new variable $k$ varies between
$K(q_{c(l,l')}^{(-)})$ and $K(q_{c(l,l')}^{(+)})$
and the variable $v$ runs from
$S(q_{s(l,l')}^{(-)})$ to $S(q_{s(l,l')}^{(+)})$.

Combining Eqs. (A3),(A4) and (A31),(A32) we find that
$\rho_{c(l,l')}(k)$ and $\rho_{s(l,l')}(v)$ are the solutions
of the following system of coupled integral
equations

\begin{equation}
2\pi\rho_{c(l,l')}(k) = 1 + {1\over {\pi}}{\cos k\over {u}}
\int_{S(q_{s(l,l')}^{(-)})}^{S(q_{s(l,l')}^{(+)})}
dv'\tilde{N}_{s(l,l')}(v'){2\pi\rho_{s(l,l')}(v') \over {1 + [v' -
{\sin k\over {u}}]^2}}
\, ,
\end{equation}
and

\begin{eqnarray}
2\pi\rho_{s(l,l')}(v) & = & {1\over {\pi}}
\int_{K_{l,l'}(q_{c(l,l')}^{(-)})}^{K(q_{c(l,l')}^{(+)})}dk'
\tilde{N}_{c(l,l')}(k'){2\pi\rho_{s(l,l')}(k') \over {1 + [v -
{\sin k'\over {u}}]^2}}\nonumber \\
& - & {1\over {2\pi}}
\int_{S_{l,l'}(q_{s(l,l')}^{(-)})}^{S(q_{s(l,l')}^{(+)})}
dv'\tilde{N}_{s(l,l')}(v'){2\pi\rho_{s(l,l')}(v') \over
{1 + [{1\over 2}(v - v')]^2}} \, ,
\end{eqnarray}
where $\tilde{N}_{c(l,l')}(k)$ and
$\tilde{N}_{s(l,l')}(v)$ are the representation of
the distributions $N_{c(l,l')}(q)$ and
$N_{s(l,l')}(q)$ (A5), respectively,
in the $k,v$ space associated with
the transformation $q\rightarrow k$ [for
$\alpha =c(l,l')$] and $q\rightarrow v$ [for
$\alpha =s(l,l')$].

For $(l,l')=(-1,-1)$ Eqs. (A33) and (A34) are similar to
Eqs. (A4) and (A5), respectively, of Ref.
\cite{Carmelo91b}. However, Eqs. (A33) and (A34) are
{\it more general}: the limits of integration of Eqs. (A4)
and (A5) of Ref. \cite{Carmelo91b} are only valid
for Hamiltonian eigenstates differing by
a small density of excited pseudoparticles
from $(18)$ (they are also valid for all
eigenstates $(15)$ where the occupied pseudomomenta
of the functions (A5) are distributed
symmetrically around the origin), whereas
Eqs. (A33) and (A34) are valid in each canonical ensemble
for all states I $(15)$ in number of $(17)$.

For the particular case of a $(-1,-1)$ state $(18)$ the
functions (A31) and (A32) are nothing but the
usual distributions of Lieb and Wu
\cite{Lieb,Carmelo91b}. Equations (A33) and
(A34) give the generalization for all states I
$(15)$.

Following Eqs. (A6), (A31), and (A32)
the distributions $\rho_{c(l,l')}(k)$ and $\rho_{s(l,l')}(v)$
obey the normaliztion conditions

\begin{equation}
{N_a\over {2\pi}}
\int_{K_{l,l'}(q_{c (l,l')}^{(-)})}^{K_{l,l'}(q_{c (l,l')}^{(+)})}
dk2\pi\rho_{c (l,l')}(k) = N^*_{c (l,l')} \, ,
\end{equation}

\begin{equation}
{N_a\over {2\pi}}
\int_{K_{l,l'}(q_{c(l,l')}^{(-)})}^{K_{l,l'}(q_{c(l,l')}^{(+)})}
dk2\pi\rho_{c(l,l')}(k)\tilde{N}_{c(l,l')}(k) = N_{c(l,l')} \, ,
\end{equation}

\begin{equation}
{N_a\over {2\pi}}
\int_{S_{l,l'}(q_{s (l,l')}^{(-)})}^{S_{l,l'}(q_{s (l,l')}^{(+)})}
dk2\pi\rho_{s (l,l')}(k) = N^*_{s (l,l')} \, ,
\end{equation}

\begin{equation}
{N_a\over {2\pi}}
\int_{S_{l,l'}(q_{s(l,l')}^{(-)})}^{S_{l,l'}(q_{s(l,l')}^{(+)})}
dk2\pi\rho_{s(l,l')}(k)\tilde{N}_{s(l,l')}(k) = N_{s(l,l')} \, .
\end{equation}
[Here the numbers $N^*_{\alpha (l,l')}$ and $N_{\alpha (l,l')}$
of the four $(l,l')$ sectors of symmetry $U(1)\otimes U(1)$
are given in Table 1.]

As mentioned above, the representation associated with
the distributions $\rho_{c(-1,-1)}(k)$ and
$\rho_{s(-1,-1)}(v)$ was, until recently, the most used in BA in
what concerns the description of states of
form $(18)$ and states whose distributions
of BA quantum numbers differ from $(18)$
by a vanishing density of these numbers
\cite{Lieb,Frahm,Shiba}. One of the reasons for this is that
from the mathematical point of view the integral
equations (A33) and (A34) are, for these states,
easier to handle than Eqs. (A3) and (A4).
However, while the distributions (A5) of
the latter equations are expectation values of the
operators $(26)$ and, therefore, have a clear physical
meaning, we note that none of the distributions
and functions $\rho_{c(l,l')}(k)$ and $\rho_{s(l,l')}(v)$,
$\tilde{N}_{c(l,l')}(k)$ and $\tilde{N}_{s(l,l')}(v)$,
and $\rho_{c(l,l')}(k)\tilde{N}_{c(l,l')}(k)$ and
$\rho_{s(l,l')}(v)\tilde{N}_{s(l,l')}(v)$ are expectation values
of any operator. (See, for instance, Fig. 3
of Ref. \cite{Carmelo88} where $\rho_{c(-1,-1)}(k)$
and the electronic momentum distribution are
compared for zero magnetic field.) The latter
distributions and functions are just a useful
{\it mathematical} representation in the $k,v$ space
for the pseudomomentum distributions (A5)
and eigenvalues $K_{l,l'}(q)$ and $S_{l,l'}(q)$ of the Hamiltonian
eigenstates $(15)$.

Finally, the expressions of the pseudoparticle bands
$(34)-(36)$ involve a third representation
for the two-pseudoparticle phase shifts (A18)-(A21)
and (A24)-(A27) in terms of the $k,v$ space variables.
The corresponding phase shifts are given by

\begin{equation}
\tilde{\Phi}_{cc}^{l,l'}(k,k') =
\bar{\Phi }_{cc}^{l,l'}\left({\sin k\over u},
{\sin k'\over u}\right) \, ,
\end{equation}

\begin{equation}
\tilde{\Phi}_{cs}^{l,l'}(k,v') =
\bar{\Phi }_{cs}^{l,l'}\left({\sin k\over u},
v'\right) \, ,
\end{equation}

\begin{equation}
\tilde{\Phi}_{sc}^{l,l'}(v,k') =
\bar{\Phi }_{sc}^{l,l'}\left(v,
{\sin k'\over u}\right) \, ,
\end{equation}
and

\begin{equation}
\tilde{\Phi}_{ss}^{l,l'}(v,v') =
\bar{\Phi }_{ss}^{l,l'}\left(v,v'\right) \, ,
\end{equation}
where the functions $\bar{\Phi }_{\alpha\alpha '}^{l,l'}$ are
defined by Eqs. (A24)-(A27).

In Appendix B we study the energy of the Hamiltonian
eigenstates I associated with the BA Eqs. (A3)-(A4) and
(A33)-(A34).

\vfill
\eject
\appendix{GENERAL ENERGY EXPRESSION FOR THE STATES I}

In this Appendix we present and study the general
energy expression for the states I with common eigenvalues
$\eta_z$ and $S_z$ corresponding to the $(l,l')$
sectors of symmetry $U(1)\otimes U(1)$.

The Hamiltonian eigenstates $(15)$ are also
eigenstates of the rapidity operator $\hat{K}_{l,l'}(q)$ and
of the operators $(26)$. Their energies can be written as

\begin{equation}
E = E_{SO(4)} - |\mu | [N^*_{c(l,l')} - N_{c(l,l')}] - \mu_0 |H|
[N^*_{s(l,l')} - N_{s(l,l')}] \, ,
\end{equation}
where

\begin{equation}
E_{SO(4)} = {N_a\over {2\pi}}
\int_{q_{c(l,l')}^{(-)}}^{q_{c(l,l')}^{(+)}}dq
\left[N_{c(l,l')}(q)\{-2t\cos K_{l,l'}(q)
- U/2\} + U/4\right] \, .
\end{equation}
Here $N_{c(l,l')}(q)$ is the $c(l,l')$ pseudomomentum
distribution (A5) and $K_{l,l'}(q)$ is
the eigenvalue of the rapidity operator
$\hat{K}_{l,l'}(q)$ defined by Eqs. (A3) and (A4): the
rapidity eigenvalues of the states I of form $(15)$,
$|\eta_z,S_z\rangle$, are fully determined by the distributions
(A5) through the BA equations (A3) and (A4). (In these
equations $S_{l,l'}(q)$ is the
eigenvalue of the rapidity operator $\hat{S}_{l,l'}(q)$.)

For the case of the state $(18)$ we denote the pseudomomentum
distribution $(26)$ by $N_{\alpha (l,l')}^0(q)$.
It is given by

\begin{eqnarray}
N_{\alpha (l,l')}^0(q) & = &
\langle 0,\eta_z,S_z|\hat{N}_{\alpha (l,l')}(q)
|0,\eta_z,S_z\rangle\nonumber \\
& = & \Theta (q_{F\alpha (l,l')}^{(+)}-q) \, ,
\hspace{0.5cm} 0<q<q_{\alpha (l,l')}^{(+)}\nonumber \\
& = & \Theta (q-q_{F\alpha (l,l')}^{(-)}) \, ,
\hspace{0.5cm} q_{\alpha (l,l')}^{(-)}<q<0 \, .
\end{eqnarray}

It is useful to express the energy (B2) in terms
of the distribution $\rho_{c (l,l')}(k)$ of Eqs. (A33)
and (A34). The term (B2) of the energy (B1) can be
rewritten in terms of that function as follows

\begin{equation}
E_{SO(4)} = {N_a\over {2\pi}}
\int_{K_{l,l'}(q_{c(l,l')}^{(-)})}^{K_{l,l'}(q_{c(l,l')}^{(+)})}
dk2\pi\rho_{c(l,l')}(k)\left[\tilde{N}_{c(l,l')}(k)\{-2t\cos k
- U/2\} + U/4\right] \, ,
\end{equation}
where $\tilde{N}_{c(l,l')}(k)$ is the distribution of Eqs.
(A33) and (A34). The two energy expressions (B1) with
$E_{SO(4)}$ given by (B2) and (B4), respectively, are equivalent.
For the sake of clarity we have here expressed the term
$E_{SO(4)}$ of the energy (B1) in both the forms
(B2) and (B4). These energies can be obtained by solving
numerically Eqs. (A3), (A4), and (B2) or Eqs. (A33),
(A34), and (B4).

As mentioned in Sec. III, in each canonical ensemble
of eigenvalues $\eta_z $ and $S_z $
the state I of maximal energy corresponds to the pseudomomentum
distribution (A5) of the particular form

\begin{eqnarray}
N_{\alpha (l,l')}^*(q) & = &
\langle *,\eta_z,S_z|\hat{N}_{\alpha (l,l')}(q)
|*,\eta_z,S_z\rangle\nonumber \\
& = & 1 - \Theta (q_{*\alpha (l,l')}^{(+)}-q) \, ,
\hspace{0.5cm} 0<q<q_{\alpha (l,l')}^{(+)}\nonumber \\
& = & 1 - \Theta (q-q_{*\alpha (l,l')}^{(-)}) \, ,
\hspace{0.5cm} q_{\alpha (l,l')}^{(-)}<q<0 \, ,
\end{eqnarray}
where

\begin{equation}
q_{*\alpha (l,l')}^{(\pm)} = q_{\alpha (l,l')}^{(\pm)} -
q_{F\alpha (l,l')}^{(\pm)} \, .
\end{equation}
Except for terms of $1/N_a$ order, the
pseudomomenta (B6) can be written as

\begin{equation}
q_{*\alpha (l,l')}^{(\pm)} \approx \pm q_{*\alpha (l,l')} \, ,
\end{equation}
where

\begin{equation}
q_{*\alpha (l,l')} = q_{\alpha (l,l')} - q_{F\alpha (l,l')} =
{\pi\over {N_a}}[N_{\alpha (l,l')}^*-N_{\alpha (l,l')}] \, .
\end{equation}
It follows that

\begin{equation}
q_{*c (l,l')} = q_{c (l,l')} - q_{Fc (l,l')} \, , \hspace{1cm}
q_{*s (l,l')} = q_{s (l,l')} - q_{Fs (l,l')} \, .
\end{equation}

The state I $|*,\eta_z,S_z\rangle$ of the rhs of Eq. (B5) is
of the form

\begin{equation}
|*;\eta_z,S_z\rangle = \prod_{\alpha=c,s}
[\prod_{q=q_{\alpha (l,l')}^{(-)}}^{q_{*\alpha (l,l')}^{(-)}}
\prod_{q=q_{*\alpha (l,l')}^{(+)}}^{q_{\alpha (l,l')}^{(+)}}
b^{\dag }_{q\alpha (l,l')}]
|V;l,l'\rangle \, .
\end{equation}

Inserting the distributions (B3) and (B5)
in Eqs. (A3) and (A4) defines the rapidities
$K^{(0)}_{l,l'}(q)$, $S^{(0)}_{l,l'}(q)$ and
$K^{(*)}_{l,l'}(q)$, $S^{(*)}_{l,l'}(q)$,
respectively. (The rapidity functions
$K^{(0)}_{l,l'}(q)$ and $S^{(0)}_{l,l'}(q)$
are also defined by Eq. $(36)$.) We call the corresponding
energies

\begin{equation}
E_0 = E^0_{SO(4)} - |\mu | [N^*_{c(l,l')} - N_{c(l,l')}] -
\mu_0 |H| [N^*_{s(l,l')} - N_{s(l,l')}] \, ,
\end{equation}
and

\begin{equation}
E_* = E^*_{SO(4)} - |\mu | [N^*_{c(l,l')} - N_{c(l,l')}] -
\mu_0 |H| [N^*_{s(l,l')} - N_{s(l,l')}] \, ,
\end{equation}
respectively, where $E^0_{SO(4)}$ and $E^*_{SO(4)}$
are given by (B2) with the $c(l,l')$ pseudomomentum
distribution defined by (B3) and (B5), respectively,
and the rapidities given by $K^{(0)}_{l,l'}(q)$ and
$K^{(*)}_{l,l'}(q)$, respectively.

The energies $E^0_{SO(4)}$ and $E^*_{SO(4)}$
can also be written in the form (B4). Let us
introduce the corresponding functions
$\rho_{c(l,l')}^0(k)$, $\rho_{s(l,l')}^0(v)$ and
$\rho_{c(l,l')}^*(k)$, $\rho_{s(l,l')}^*(v)$, respectively.
These are solutions of the integral equations
(A33) and (A34) with

\begin{equation}
\tilde{N}_{c(l,l')}^0(k) = \Theta (Q_{l,l'}-|k|) \, , \hspace{1cm}
\tilde{N}_{s(l,l')}^0(v) = \Theta (B_{l,l'}/u-|v|) \, ,
\end{equation}
and

\begin{equation}
\tilde{N}_{c(l,l')}^*(k) = 1 - \Theta (Q_{l,l'}^*-|k|) \, , \hspace{1cm}
\tilde{N}_{s(l,l')}^*(v) = 1 - \Theta (B_{l,l'}^*/u-|v|) \, ,
\end{equation}
respectively, where $Q_{l,l'}$ and $B_{l,l'}$ are defined
in Eqs. (A22) and

\begin{equation}
Q_{l,l'}^* = K^{(*)}_{l,l'}(q_{*c (l,l')}) \, ,
\hspace{1cm}
B_{l,l'}^*/u = S^{(*)}_{l,l'}(q_{*s (l,l')}) \, ,
\end{equation}
respectively. In Eqs. (A22) and (B13)-(B15) we
have used the values $(22)$ and (B7)-(B9). For both the
states $(18)$ and (B10) the limits of integration of
Eqs. (A33) and (A34) are given by

\begin{equation}
K^{(0)}_{l,l'}(\pm q_{c(l,l')}) =
K^{(*)}_{l,l'}(\pm q_{c(l,l')}) = \pm \pi \, ,
\hspace{1cm}
S^{(0)}_{l,l'}(\pm q_{s(l,l')}) =
S^{(*)}_{l,l'}(\pm q_{s(l,l')}) = \pm \infty \, .
\end{equation}

\vfill
\eject
\appendix{STATE OF MINIMAL ENERGY OF A FAMILY OF STATES\\
          WITH COMMON ETA SPIN AND SPIN}

In this Appendix we show that at fixed and finite values
of the chemical potential and magnetic field
the lowest-energy Hamiltonian eigenstate of
a family of states $|\eta ,S; \eta_z , S_z\rangle $
with common values $\eta$ and $S$ but different
eigenvalues $\eta_z$ and $S_z$ is the
[LWS,LWS], [LWS,HWS], [HWS,LWS], or
[HWS,HWS] corresponding to the $(l,l')$
sector choosed by the signs of the chemical
potential and magnetic field.

Obviously, we have that

\begin{equation}
\hat{H}_{SO(4)}|\eta ,S; \eta_z , S_z\rangle =
E_{SO(4)}(\eta ,S)|\eta ,S; \eta_z , S_z\rangle \, ,
\end{equation}
where $E_{SO(4)}(\eta ,S)$ is the eigenenergy of the whole
family of states of eta spin $\eta$ and spin $S$
corresponding to the Hamiltonian $(2)$. Moreover,
the eigenenergy of $|\eta ,S; \eta_z , S_z\rangle$
relative to the Hamiltonian $(1)$ is

\begin{equation}
E(\eta_z , S_z) = E_{SO(4)}(\eta ,S) +
2\mu {\eta }_z + 2\mu_0 HS_z \, ,
\end{equation}
whereas the eigenenergy of the four above LWS's and (or)
HWS's is

\begin{equation}
E_{[LWS,LWS]} = E_{SO(4)}(\eta ,S) -
2\mu\eta - 2\mu_0 HS \, ,
\end{equation}

\begin{equation}
E_{[LWS,HWS]} = E_{SO(4)}(\eta ,S) -
2\mu\eta + 2\mu_0 HS \, ,
\end{equation}

\begin{equation}
E_{[HWS,LWS]} = E_{SO(4)}(\eta ,S) +
2\mu\eta - 2\mu_0 HS \, ,
\end{equation}
and

\begin{equation}
E_{[HWS,HWS]} = E_{SO(4)}(\eta ,S) +
2\mu\eta + 2\mu_0 HS \, .
\end{equation}
For $\eta_z\neq \pm\eta$ and (or) $S_z\neq \pm S$
out of the four energies

\begin{equation}
E(\eta_z , S_z) - E_{[LWS,LWS]} =
2\mu (\eta_z + \eta ) + 2\mu_0 H (S_z + S) \, ,
\end{equation}

\begin{equation}
E(\eta_z , S_z) - E_{[LWS,HWS]} =
2\mu (\eta_z + \eta ) + 2\mu_0 H (S_z - S) \, ,
\end{equation}

\begin{equation}
E(\eta_z , S_z) - E_{[HWS,LWS]} =
2\mu (\eta_z - \eta ) + 2\mu_0 H (S_z + S) \, ,
\end{equation}
and

\begin{equation}
E(\eta_z , S_z) - E_{[HWS,HWS]} =
2\mu (\eta_z - \eta ) + 2\mu_0 H (S_z - S) \, ,
\end{equation}
the energy difference corresponding to the state
$|\eta ,S; -\eta , -S\rangle $, $|\eta ,S; -\eta , S\rangle $,
$|\eta ,S; \eta , -S\rangle $, or $|\eta ,S; \eta , S\rangle $
which belongs to the $(l,l')$ sector choosed by the
signs of the chemical potential $\mu$ and magnetic
field $H$ is positive and maximal. This follows
directly from the form of the four energies
(C7)-(C10).

For example, consider that $\mu >0$ and $H>0$. In this
case the eta-spin and spin projections are such that
$\eta_z<0$ and $S_z<0$ [see Eq. $(29)$], respectively, and we are
in the $(-1,-1)$ sector of symmetry $U(1)\otimes U(1)$. It follows
then from Eqs. (C7)-(C10) that the [LWS,LWS] Hamiltonian
eigenstate $|\eta ,S; -\eta , -S\rangle $ has minimal energy
and the gap $E(\eta_z , S_z) - E_{[LWS,LWS]}$ is
positive when $\eta_z\neq -\eta$ and (or) $S_z\neq -S$.
(Also the energies $E_{[LWS,HWS]} - E_{[LWS,LWS]}$,
$E_{[HWS,LWS]} - E_{[LWS,LWS]}$, and
$E_{[HWS,HWS]} - E_{[LWS,LWS]}$ are in this case
positive.) There are two minimal gaps, $\Delta_{\eta}$
and $\Delta_S$, which correspond to eta spin and spin,
respectively. They refer to the states
$|\eta ,S; -\eta + 1, -S\rangle $ and
$|\eta ,S; -\eta , -S+1\rangle $, respectively.
Following Eq. (C7) they read $\Delta_{\eta} = 2\mu$
and $\Delta_{S} = 2\mu_0 H$, respectively.

The same applies for other signs of the chemical
potential and magnetic field, i.e. for the remaining
three $U(1)\otimes U(1)$ sectors. The corresponding LWS and
(or) HWS is the state of minimal energy of the family of
states $|\eta ,S; \eta_z , S_z\rangle$. In the general case,
the minimal gaps are given by

\begin{equation}
\Delta_{\eta} = 2|\mu| \, ;\hspace{2cm} \Delta_{S} =
2\mu_0 |H| \, .
\end{equation}


\newpage
\narrowtext
\begin{tabbing}
\sl \hspace{2cm} \= \sl $(-1,-1)$ \hspace{0.3cm} \=
\sl $(-1,1)$ \hspace{0.3cm} \= \sl $(1,-1)$
\hspace{0.7cm} \= \sl $(1,1)$\\
$N_{c(l,l')}$ \> $N$ \> $N$ \> $2N_a-N$ \> $2N_a-N$\\
$N_{c(l,l')}^*$ \> $N_a$ \> $N_a$ \> $N_a$ \> $N_a$\\
$N_{s(l,l')}$ \> $N_{\downarrow}$ \> $N_{\uparrow}$
\> $N_a-N_{\uparrow}$ \> $N_a-N_{\downarrow}$\\
$N_{s(l,l')}^*$ \> $N_{\uparrow}$ \> $N_{\downarrow}$
\> $N_a-N_{\downarrow}$ \> $N_a-N_{\uparrow}$\\
$q_{Fc(l,l')}$ \> $2k_F$ \> $2k_F$ \> $2[\pi-k_F]$ \> $2[\pi-k_F]$\\
$q_{c(l,l')}$ \> $\pi$ \> $\pi$ \> $\pi$ \> $\pi$\\
$q_{Fs(l,l')}$ \> $k_{F\downarrow}$ \> $k_{F\uparrow}$
\> $\pi-k_{F\uparrow}$ \> $\pi-k_{F\downarrow}$\\
$q_{s(l,l')}$ \> $k_{F\uparrow}$ \> $k_{F\downarrow}$
\> $\pi-k_{F\downarrow}$ \> $\pi-k_{F\uparrow}$
\label{table1}
\end{tabbing}
\vspace{0.5cm}
TABLE 1 -- Values of the numbers $N_{\alpha (l,l')}$
and $N_{\alpha (l,l')}^*$ and of the pseudo-Fermi
points and limits of the pseudo-Brillouin zones $(22)$ in
the four $(l,l')$ sectors of symmetry $U(1)\otimes U(1)$.
\end{document}